\documentclass[12pt]{article}
\usepackage{amsmath,amssymb,amsfonts,graphicx,slashed,color,cite}
\usepackage[colorlinks=true,linkcolor=blue, citecolor=blue, linktoc=all]{hyper ref}

\DeclareMathAlphabet{\mathbbold}{U}{bbold}{m}{n}

\def\uiucaddress{\small Department of Physics, University of Illinois, 1110 W. Green St., 
Urbana IL 61801-3080, U.S.A. }
\def\title{\LARGE{Modular Hamiltonians for Deformed Half-Spaces and the Averaged Null Energy Condition
}}

\parskip=\baselineskip

\newcommand{\myfig}[3]{
	\begin{figure}[ht]
	\centering
	\includegraphics[width=#2cm]{#1}\caption{#3}\label{fig:#1}
	\end{figure}
	}

\newcommand{\cDsl}{{{\cal D}\kern-.65em /\,}}
\newcommand{\cDslsm}{{{\cal D}\kern-.5em /\,}}
\newcommand{\nabslsm}{\nabla\kern-.55em /}
\newcommand{\pasl}{\pa\kern-.55em /}
\newcommand{\psl}{p\kern-.45em /}
\newcommand{\Dsl}{D\kern-.65em /}
\newcommand{\Asl}{A\kern-.55em /}
\newcommand{\nabsl}{\nabla\kern-.65em /\kern+.2em}
\newcommand{\qsl}{q\kern-.5em /}
\newcommand{\ksl}{k\kern-.5em /}
\newcommand{\rsl}{r\kern-.5em /}

\newcommand{\cDslLCsq}{{\stackrel{\circ}{\cDsl^{\kern2pt 2}}}}

\newcommand{\cL}{\mathcal{L}}

\newcommand\cc[1]{#1^{^{\kern-6pt \circ}}\kern2pt}

\renewcommand{\a}{\alpha}
\renewcommand{\b}{\beta}

\newcommand{\re}{\mathbb{R}}

\newcommand{\bx}{{\mathbf{x}}}

\newcommand{\pa}{\partial}

\newcommand{\beq}{\begin{equation}}
\newcommand{\eeq}{\end{equation}}
\newcommand{\beqn}{\begin{eqnarray}}
\newcommand{\eeqn}{\end{eqnarray}}

\def\dalemb#1#2{{\vbox{\hrule height .#2pt
\hbox{\vrule width.#2pt height#1pt \kern#1pt
\vrule width.#2pt}
\hrule height.#2pt}}}

\newcommand{\reg}{{A_0}} 
\newcommand{\dreg}{A} 
\newcommand{\regc}{{A_0^c}} 
\newcommand{\dregc}{A^c} 

\newcommand{\cK}{\widehat{K}}

\newcommand{\id}{\mathbbold{1}}
\newcommand{\cO}{\mathcal{O}}

\begin{document}

\begin{center}
\title
\end{center}
\vskip 2 cm
\centerline{{\bf 
Thomas Faulkner, Robert G. Leigh, Onkar Parrikar and Huajia Wang}}

\vspace{.5cm}
\centerline{\it \uiucaddress}
\vspace{2cm}

\begin{abstract}
We study modular Hamiltonians corresponding to the vacuum state for deformed half-spaces in relativistic quantum field theories on $\re^{1,d-1}$. We show that in addition to the usual boost generator, there is a contribution to the modular Hamiltonian at first order in the shape deformation, proportional to the integral of the null components of the stress tensor along the Rindler horizon. We use this fact along with monotonicity of relative entropy to prove the averaged null energy condition 
in Minkowski space-time.  This subsequently gives a new proof of the Hofman-Maldacena bounds on the parameters appearing in CFT three-point functions.  Our main technical advance involves adapting newly developed perturbative methods for calculating entanglement entropy to the problem at hand. These methods were recently used to prove certain results on the shape dependence of entanglement in CFTs and here we generalize these results to excited states and real time dynamics.  We also discuss the AdS/CFT counterpart of this result, making connection with the recently proposed gravitational dual for modular Hamiltonians in holographic theories.
\end{abstract}

\pagebreak

\section{Introduction}
The entanglement structure of states is of great importance in quantum field theory. The most common tool used for studying entanglement structure is the entanglement entropy, namely the Von Neumann entropy of a subregion, and has already provided many important insights. A more fine-grained probe is the modular Hamiltonian, defined as 
\beq
K_A^{\Psi} = - \mathrm{ln}\, \rho_{A}^{\Psi}
\eeq
where $\rho^{\Psi}_A$ is the reduced density matrix of the state $\Psi$ over the subregion $A$. The modular Hamiltonian is, in general, a complicated, non-local operator and not of much practical use. However, the situation greatly simplifies for the vacuum state in the case of certain special symmetric subregions. For instance, the modular Hamiltonian for a half-space in relativistic quantum field theories takes a very simple form; it is the restriction to the half space of the generator of boosts which preserve the entangling surface \cite{Bisognano:1976za}, and consequently generates a local and geometric modular flow. A similar construction is also possible for spherical subregions in conformal field theories \cite{Casini:2011kv}, for null slabs in the case where the vacuum state is defined with respect to the generator of null translations on a null hypersurface \cite{Bousso:2014sda, Bousso:2014uxa} etc. Recently, it has been argued that the modular Hamiltonian for states with classical gravitational duals also takes a simple form \cite{Jafferis:2014lza, Jafferis:2015del}. On the other hand, it is quite non-trivial to study the entanglement entropy and modular Hamiltonian for more general (asymmetric) subregions, especially outside the purview of free-field theories or $AdS$/CFT. Some progress was made in \cite{Faulkner:2015csl} (following previous work in \cite{Rosenhaus:2014woa, Rosenhaus:2014zza, Allais:2014ata, Mezei:2014zla, Faulkner:2014jva}), where perturbative techniques were used to study the shape-dependence of entanglement entropy in conformal field theories. In the present paper, we adapt these techniques to study modular Hamiltonians for deformed half-spaces in relativistic (not necessarily conformal) quantum field theories.

The study of shape dependence of entanglement is an important task for several reasons. The entanglement structure of quantum systems is highly constrained by powerful inequalities, such as strong subadditivity of entanglement entropy, positivity and monotonicity of relative entropy, etc. In many situations, these entanglement inequalities further imply fundamental constraints on the properties of quantum field theories. For instance, the strong subadditivity property was used in \cite{Casini:2004bw,Casini:2012ei} to prove an entropic version of the c-theorem for renormalization group flows in two and three dimensions. Similarly, the properties of relative entropy have been used to prove several interesting results such as the Bekenstein bound \cite{Casini:2008cr}, the generalized second law for causal horizons \cite{Wall:2011hj} and the covariant entropy bound in the context of semi-classical gravity \cite{Bousso:2014sda, Bousso:2014uxa}. Entanglement inequalities have also been shown to constrain the bulk geometry in states with classical gravity duals \cite{Lashkari:2014kda, Lashkari:2016idm, Banerjee:2014oaa, Bhattacharyya:2016knk}. The entanglement inequality which will be relevant for our purpose is that the \emph{full} modular Hamiltonian for the \emph{vacuum} state
\beq
\cK_A = K_A - K_{A^c}
\eeq
(i.e., the difference between the modular Hamiltonian of the subregion $A$ and that of the complementary subregion $A^c$) satisfies a ``monotonicity'' property under inclusion \cite{zbMATH00845667}. This means that if we shrink the subregion $A$, then the corresponding change in the full modular Hamiltonian $\delta \cK_A$ is a negative semi-definite operator. This property in fact follows from the monotonicity of relative entropy, as was shown in \cite{Blanco:2013lea}. In the present work, we will show that this monotonicity property of the full modular Hamiltonian along with perturbative results on the shape dependence of the modular Hamiltonian allow us to prove another fundamental constraint, namely the \emph{averaged null energy condition} (ANEC)
\beq
\int_{-\infty}^{\infty} dx^+ \left\langle T_{++} (x^+,x^-=0,\vec{x}^i) \right\rangle_{\psi} \geq 0 
\eeq
for excited state in relativistic quantum field theories on Minkowski space-time. 

From a classical general relativity point of view, averaged energy conditions (which are weaker than the point-wise null, weak, strong, dominant-energy conditions) were shown to be sufficient for proving a number of interesting results such as standard focussing theorems \cite{Borde:1987qr, Tipler:1978zz}, topological censorship \cite{Friedman:1993ty}, etc. This provides a clear motivation for trying to prove or disprove averaged energy conditions for \emph{quantum} fields in general space-times, given that most point-wise conditions are known to be violated by quantum effects (see for instance, \cite{Fewster:2012yh}).\footnote{However, there are alternative proposals for point-wise quantum energy conditions. See for example \cite{Bousso:2015mna, Bousso:2015wca, Koeller:2015qmn}.} In Minkowski space-time, the ANEC has been proven to hold for many special cases such as free scalar and Maxwell fields in general dimension \cite{Klinkhammer:1991ki, Ford:1994bj, Folacci:1992xg}, arbitrary quantum field theories in $d=2$ with a mass gap and some assumptions on the stress tensor \cite{Verch:1999nt}, CFTs with classical gravitational duals in general dimension \cite{Kelly:2014mra}, etc. Wall has also argued that the ANEC holds true for free or superrenormalizable field theories in general dimension \cite{Wall:2011hj}. In summary, there is substantial evidence so far to suggest that the ANEC is satisfied by generic quantum field theories on Minkowski space-time, but a general proof has been missing hitherto (although see \cite{Hofman:2009ug} for an argument involving certain assumptions on the OPE of non-local operators) -- in this paper, we will partially fill this gap. On the other hand, the ANEC is known to be violated in general curved space-times, but an alternative proposal called the self-consistent achronal ANEC exists in this case -- see \cite{Flanagan:1996gw, Graham:2007va, Kontou:2015yha} and references there-in for further discussion. While this is out of the scope of the present paper, our results can nevertheless be extended to prove the ANEC along static bifurcate Killing horizons even in curved space-times.  

There is another motivation for trying to prove the ANEC in Minkowski space-time. In \cite{Hofman:2008ar}, Hofman and Maldacena (HM) showed that in a conformal field theory the validity of the ANEC in a certain class of states created by operator insertions implies bounds on the coefficients appearing in the three-point correlation functions of that CFT. For instance, in $d=4$ they used this to derive a bound on the ratio of central charges
\beq
\frac{1}{3} \leq \frac{a}{c} \leq \frac{31}{18}
\eeq
where $a$ and $c$ are the coefficients of the Euler density term and the Weyl tensor squared term in the conformal anomaly; tighter bounds can be obtained by imposing supersymmetry. 
While the assumption of the ANEC was considered reasonable, in the original paper no proof was given. Since then, there have been several attempts at a proof of the HM bounds with varying levels of success \cite{Hofman:2009ug, Kulaxizi:2010jt, Komargodski:2016gci}.

In particular, using analytic bootstrap methods the HM bounds were proven for a class of three-point functions in \cite{Hofman:2016awc}, building on the work of \cite{Hartman:2016dxc, Hartman:2015lfa}.  These methods take as an input crossing symmetry and reflection positivity and apply these principles to various four point functions in a light-cone  limit to delicately extract the HM bounds. In particular in this guise the HM bounds were related to causality properties of correlation functions in a shockwave background \cite{ Hartman:2015lfa}.\footnote{In theories with gravity duals, the HM bounds have also been shown to be related to bulk causality constraints \cite{Hofman:2009ug, Camanho:2009vw, Buchel:2009sk, Kelly:2014mra}.} In contrast, we will show that the general HM constraints on CFT three-point functions can be extracted directly from the three-point function itself - when the three-point function is interpreted as calculating some modular energy of the CFT in an excited state.

Overall it is satisfying to see the ANEC, and consequently the Hofman-Maldacena bounds, arise as a natural consequence of the fundamental constraints satisfied by the entanglement structure of the vacuum.


\subsection{Setup \&\ Summary of results}
We now outline the calculation we are interested in, and present a brief summary of our results. Consider the density matrix $|\Psi\rangle\langle\Psi|$ corresponding to a pure state defined on the Cauchy surface $\Sigma$. Let us partition $\Sigma$ into two subregions $A$ and its complement $A^c$. For local quantum field theories, we expect the Hilbert space $\mathfrak{h}_{\Sigma}$ to factorize into the tensor product $\mathfrak{h}_{\Sigma}=\mathfrak{h}_{A}\otimes \mathfrak{h}_{A^c}$. If this is the case, we can trace over $\mathfrak{h}_{A^c}$ to obtain the reduced density matrix
\beq
\rho^{\Psi}_A = \mathrm{Tr}_{A^c}(|\Psi\rangle\langle\Psi|)
\eeq 
which contains all the relevant information pertaining to the subregion $A$. The \emph{entanglement entropy}  between $A$ and $A^c$ is defined as the von Neumann entropy of $\rho^{\Psi}_A$
\beq
S_{EE}[\Psi, A] = -\mathrm{Tr}_{A}\left(\rho^{\Psi}_A\,\mathrm{ln}\;\rho^{\Psi}_A\right).
\eeq
In this context, the boundary $\pa A$ of $A$ is referred to as the \emph{entangling surface}. The \emph{modular Hamiltonian} (also known as the entanglement Hamiltonian) $K^{\Psi}_A$ is defined as
\beq
K^{\Psi}_A = -\mathrm{ln}\,\rho^{\Psi}_A
\eeq
Similarly, we can also define the modular Hamiltonian corresponding to the region $A^c$, which we denote $K_{A^c}^{\Psi}$. We can combine $K^{\Psi}_A$ and $K^{\Psi}_{A^c}$ into another useful  operator:
\beq
\cK_A^{\Psi} = K_A^{\Psi}\otimes \id_{A^c} - \id_{A}\otimes K_{A^c}^{\Psi}
\eeq
which we will refer to as the \emph{full modular Hamiltonian}.

In this paper, we will primarily study the operators $K_A,\,K_{A^c}$ and $\cK_A$ for the \emph{vacuum} state of a relativistic quantum field theory (as such we drop the label $\Psi$ from now on), with the region $A$ being a slightly deformed half-space. To specify the geometry in more detail, let us pick global coordinates $x^{\mu}=(x^0, x^1,\cdots, x^{d-1})= (x^0,\bx)$ on $\re^{1,d-1}$, where $x^0$ is the time coordinate, and $\bx$ denotes spatial coordinates. Pick the Cauchy surface $\Sigma$ given by $x^0=0$, and consider the half space $\reg$ given by
\beq
\reg =\left\{x^{\mu}\in \re^{1,d-1} | x^0=0,\;x^1 > 0\right\}.
\eeq
The vacuum modular Hamiltonian for the half space takes a particularly simple form \cite{Bisognano:1976za}
\beq
K_{\reg} = 2\pi\int_{\reg}d^{d-1}\bx \,x^1\, T_{00}(0,\bx)+\mathrm{constant}
\eeq
i.e., it is the generator of boosts which preserve the entangling surface restricted to the region $\reg$, a result known as the Bisognano-Wichmann theorem \cite{Bisognano:1976za}. Correspondingly, the full modular Hamiltonian is given by the full boost generator
\beq
\cK_{\reg} = 2\pi\int_{\Sigma}d^{d-1}\bx \,x^1\, T_{00}(0,\bx)
\eeq
Note that $\cK_{\reg}$ is a conserved charge, and as such annihilates the vacuum $\cK_{\reg}|0\rangle= 0$. (This later property is true for more general regions as well.) 

\myfig{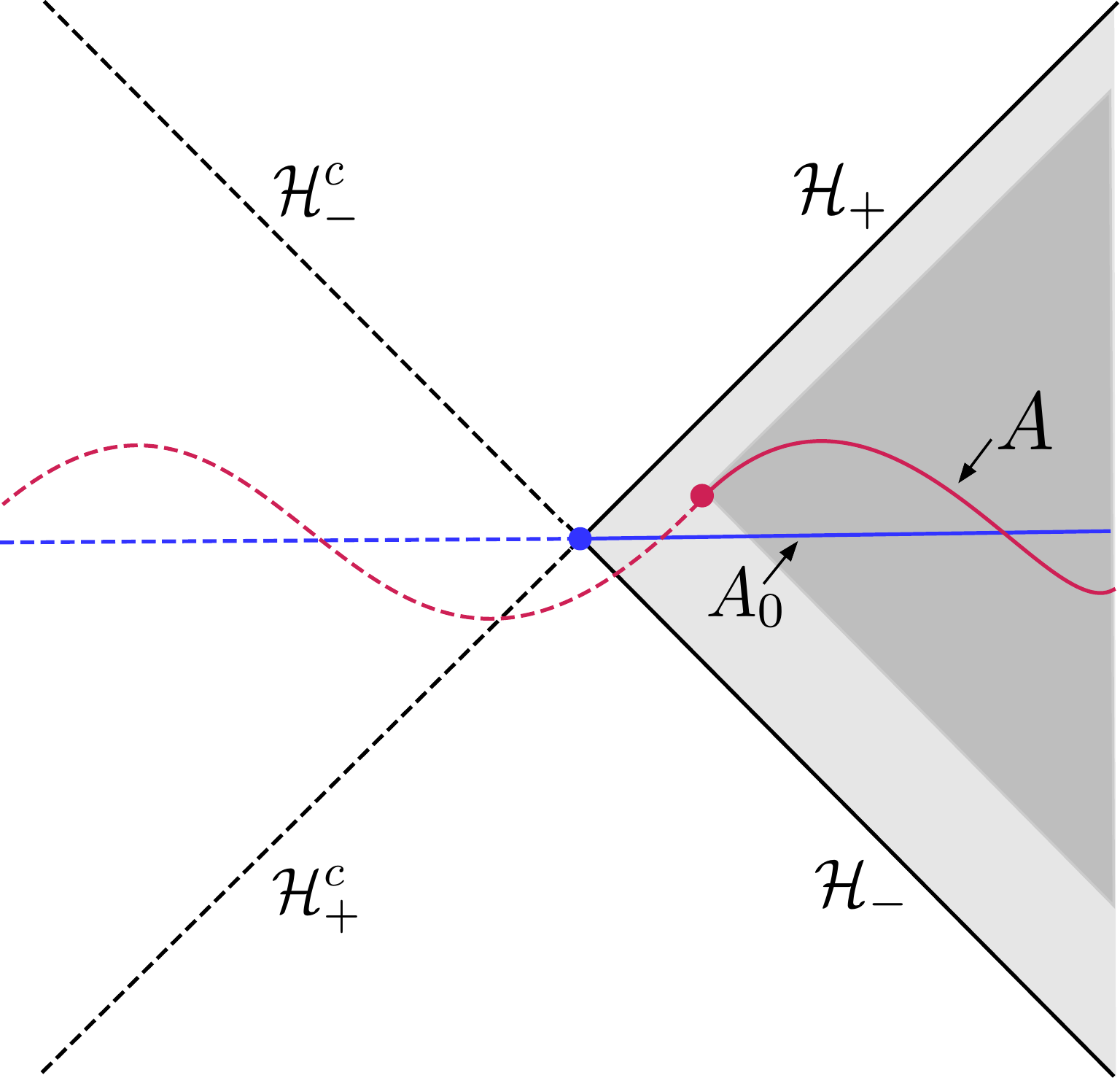}{7.2}{\small{\textsf{We deform the half space $\reg$ (solid blue line) inwards into the region $\dreg$ (solid red line), such that $\mathcal{D}(\dreg)$ (darker shaded region) is contained inside $\mathcal{D}(\reg)$ (lighter shaded region). Also shown are the Rindler horizons $\mathcal{H}_{\pm}$ corresponding to the regions $\reg$ and $\regc$. (The transverse directions $\vec{x}$ are implicit.)}}}

One can consider a small deformation of the region $\reg$ to 
\beq
\dreg = \left\{x^{\mu}\in \re^{1,d-1}\ \Big|\ x^0=0,\;x^1 >\zeta(\vec{x})\right\}
\eeq
where $\zeta(\vec{x})$ is a smooth function of the $(d-2)$ transverse spatial coordinates (parametrizing the entangling surface), collectively denoted by $\vec{x}=(x^2,\cdots,x^{d-1})$. The deformation is special in that it is restricted within the original Cauchy surface. We can generalize this to also include time-like deformations (see figure \ref{fig:fig1.pdf})
\beq
\dreg = \left\{\tilde{x}^{\mu}\in \re^{1,d-1}\ \Big|\ \tilde{x}^0=\zeta^0(x^1,\vec{x}),\;\tilde{x}^1=x^1+\zeta^1(x^1,\vec{x}),\;x^1>0\right\}.
\eeq
We identify the infinitesimal $\zeta^\mu$ as the deformation vector field and pick $\zeta^{\mu}$ to point inward, i.e. $\mathcal{D}(\dreg) \subset \mathcal{D}(\reg)$ (where $\mathcal{D}$ denotes the domain of dependence). 

Our primary results in this paper are as follows: 

(i) We will first show that the modular Hamiltonian $K_{\dreg}$, up to first order in the shape deformation, is given by 
\beq
UK_{\dreg}U^{\dagger} = K_{\reg} -2\pi\int_{\mathcal{H}_+}\zeta^+T_{++}+2\pi\int_{\mathcal{H}_-}\zeta^-T_{--}+2\pi \int_{\reg}  \zeta^\nu\left[ K_{\reg}, T_{0\nu} \right]+ O(\zeta^2) \label{mH}
\eeq
where $U:\mathfrak{h}_{\dreg}\to \mathfrak{h}_{\reg}$ is a unitary transformation the details of which we will specify later, $\mathcal{H}_{\pm}$ are the future and past Rindler horizons of $\mathcal{D}(A_0)$ shown in figure \ref{fig:fig1.pdf}, and $\zeta^{\pm}$ are components of the vector field $\zeta$ in light-cone coordinates $x^{\pm} = x^0\pm x^1$. Similar expressions can also be written for $K_{\dregc}$ and $\cK_{\dreg}$.

(ii) We will then consider the expectation value $\langle\psi| \cK_A|\psi\rangle $ in  states of the form
\beq\label{states}
|\psi_{\a} \rangle = e^{-\tau H} \mathcal{O}_{\a}(0,\bx)|0\rangle
\eeq
and linear combinations thereof, where $H$ is the Hamiltonian and $\mathcal{O}_{\a}$ is an arbitrary local operator whose quantum numbers (dimension, spin etc.) are collectively denoted by $\a$.\footnote{In the case of tensor operators, we contract them with appropriate polarizations, for instance $\cO_{\alpha}(x) = \epsilon_{\mu_1\mu_2\cdots \mu_s}J^{\mu_1\mu_2\cdots \mu_s}(x)$} 
The factor of $e^{-\tau H}$ is added to make these states normalizable. 
In a CFT this class of states is a basis for the entire Hilbert space, via the state-operator mapping. For a general QFT similar statements should hold. In fact, there is no obstruction to generalizing our argument to include states created by many local and even non-local operators inserted throughout the lower half Euclidean plane. Further, we could also insert the operators in real time. In the interest of simplifying our presentation we choose to represent our state via a single operator insertion on the Euclidean section, although we expect all our conclusions to go through even in the more general case.

We then show that equation \eqref{mH}, along with the positivity of the operator $\cK_{\reg}-\cK_{\dreg}$ (i.e. monotonicity under inclusion, which recall follows from the monotonicity of relative entropy) implies the averaged null energy condition (ANEC)
\beq
\int_{-\infty}^{\infty} dx^+\left\langle T_{++}(x^+,x^-=0,\vec{x})\right\rangle_{\psi} \geq 0. 
\eeq
As discussed in the introduction, the Hofman-Maldacena bounds on CFT three-point functions were derived assuming the ANEC; so this completes the proof of these bounds. It should perhaps be mentioned that the specific states considered in deriving the HM bounds in \cite{Hofman:2008ar} were created by inserting an approximately local operator in real time, with an approximately specified four-momentum. Given our remarks below equation \eqref{states}, our derivation of the ANEC also applies to these states.

(iii) Finally, we also discuss the (vacuum) full modular Hamiltonian for deformed half-spaces in CFTs with classical gravity duals, which allows us to make contact with the recent proposal by Jafferis-Lewkowyzc-Maldacena-Suh (JLMS) \cite{Jafferis:2015del} for the holographic dual to the modular Hamiltonian. 


At this point we should mention that in continuum quantum field theory there are significant ultraviolet 
(UV) issues associated with the definition of the reduced density matrix for a region, often resulting
in divergences for entanglement entropy and modular energy which are local to the entangling surface. These issues and associated divergences are however not present for quantities like the relative entropy, and the full modular Hamiltonian \cite{araki1976relative,haag2012local}.
Since this is ultimately what we are interested in, and in the interest of simplicity of presentation, we will for the most part suppress the need for a UV cutoff at the entangling surface. Indeed the answers we will find will be finite, partly justifying this approach. For further discussion on how to include such a UV cutoff in our calculation, see Appendix~\ref{app:cutoff}, where we will argue for the irrelevance of the details of such a cutoff beyond its existence. 

\section{Modular Hamiltonian for deformed half-space}
\label{sec:moddef} 
In this section, we give an explicit formula for the modular Hamiltonian $K_{A}$ of the \emph{vacuum} state over a deformed half-space, to first order in the shape deformation. 
\subsection{Reduced density matrix}
The vacuum state in a relativistic quantum field theory can be constructed by performing the Euclidean path-integral over the region $x_E^0<0$, where $x_E^0$ is Euclidean time. 
In the interest of generality, let us instead consider a more general state rather than the vacuum\footnote{Later, we will also need to compute the expectation value $\langle K_{\dreg} \rangle_\psi=\text{Tr}_A\left(\rho^\psi_A K_{A,}\right)$ in the excited state $|\psi\rangle$; so we derive the reduced density matrix $\rho^\psi_A$ along the way while setting up the calculation for $K_A$.} 
\beq
|\psi\rangle = \sum_{\a} c_{\a}|\psi_{\a}\rangle  =  e^{-\tau H}\sum_{\a} c_{\a}  \mathcal{O}_{\a}(0,\bx)|0\rangle,\;\;\;\cdots\;\; (\tau >0)
\eeq
This state can be constructed similarly as a sum over path-integrals, but with the operator $\mathcal{O}_{\a}$ inserted at $x^0_E = -\tau$ in the term proportional to $c_{\alpha}$. The reduced density matrix corresponding to $|\psi\rangle$, associated with the undeformed half-space $\reg$ is constructed as a Euclidean path integral with specified field configurations above ($x_E^0\to 0^+,\,x^1>0$) and below ($x_E^0\to 0^-,\,x^1>0$) the region $\reg$ (see Fig. \ref{fig:fig2.pdf})
\beq\label{eq:diff_rho}
\langle\alpha_0| \rho^{\psi}_{\reg,\eta} |\beta_0\rangle={\cal N}^{-1}_{\reg,\eta}\int^{\phi^+=\beta_0}_{\phi^-=\alpha_0}[D\phi]_\eta\; \sum_{\a}\sum_{\a'}c^*_{\a}c_{\a'} \cO_{\a}^{\dagger}(\tau,\bx) \cO_{\a'}(-\tau,\bx)\;e^{-S[\eta,\phi(x)]}
\eeq
where we have collectively denoted all the fields integrated over in the path integral as $\phi$, and $|\a_0\rangle, \b_0\rangle  \in \mathfrak{h}_{\reg}$ are eigenstates of the field operator $\phi$ restricted to $\reg$. The prefactor $\mathcal{N}^{-1}_{\reg,\eta}$ is added to ensure the normalization of the density matrix, i.e. $\mathrm{Tr}_{\reg}\,\rho^{\psi}_{\reg,\eta}=1$. We have explicitly displayed the dependence of the reduced density matrix on the metric $\eta_{\mu\nu}$\footnote{Here by $\eta$ we are denoting the metric in real time. Of course the corresponding metric on the Euclidean section which is used in constructing the Euclidean path integral is $\delta_{\mu\nu}$.} through the path integral measure (which we assume is diffeomorphism invariant), the action and the normalization.

For convenience, we will henceforth use the notation (inside path-integrals)
\beq\label{Oinsertions}
X=\sum_{\a}\sum_{\a'}c^*_{\a}c_{\a'} \cO_{\a}^{\dagger}(\tau,\bx) \cO_{\a'}(-\tau,\bx).
\eeq

\myfig{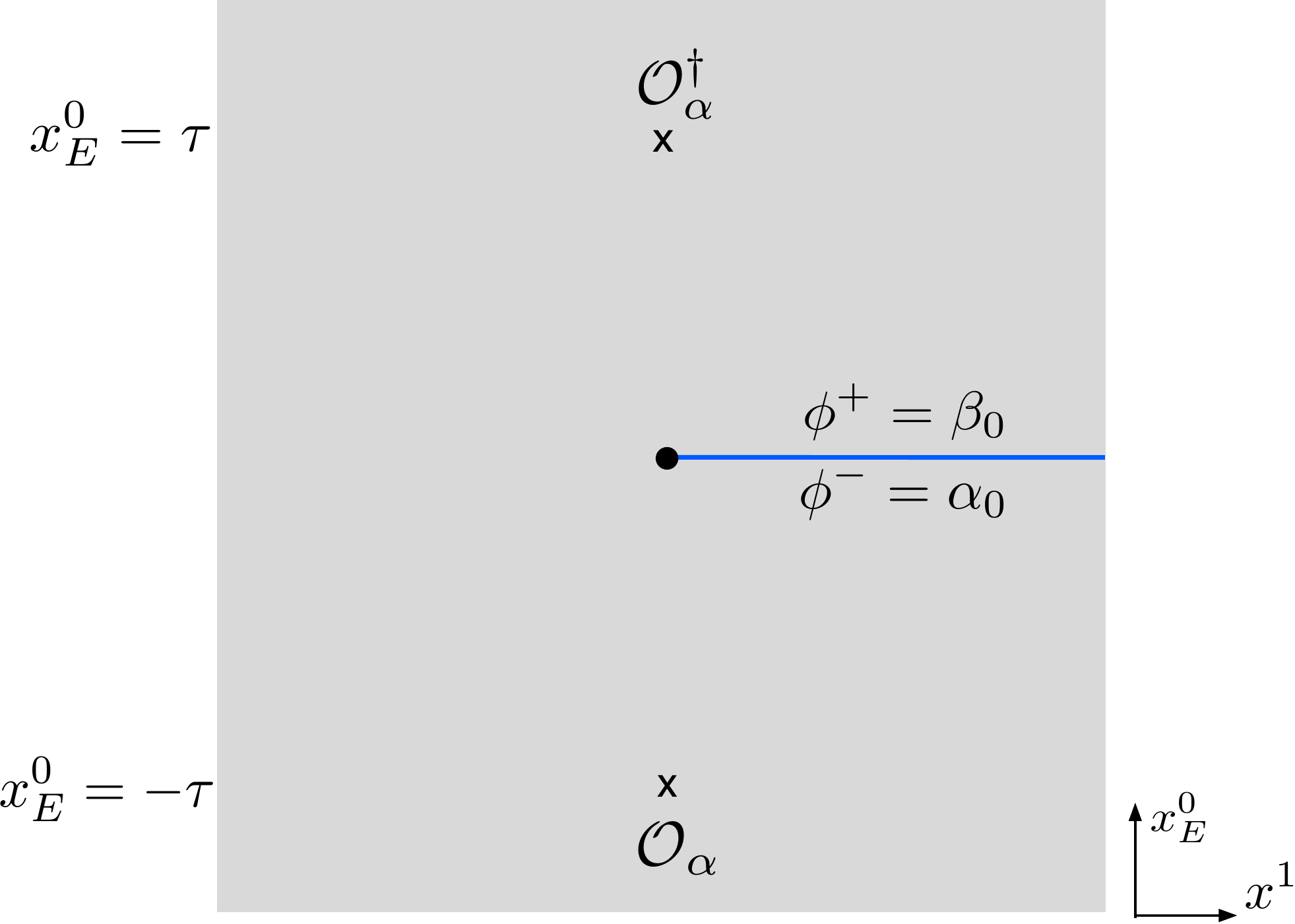}{9}{\small{\textsf{ The path integral construction for matrix elements of the reduced density matrix for the state $|\psi_{\a}\rangle$, over the original half space $\reg$ (solid blue line). The operator insertions are marked at $x_E^0 = \pm \tau$.  The black dot is the entangling surface (with transverse directions $\vec{x}$ implicit).  }} }


Now consider the reduced density matrix over the deformed region $\dreg$. This can be constructed by a similar Euclidean path integral with specified field configurations above and below $\dreg$, and with a real-time fold around $x_E^0=0$ in the case of time-like deformations ($\zeta^0\neq 0$). We can deal with this path integral by performing a diffeomorphism $f: x^\mu\to x^\mu -\zeta^\mu$, which maps $\dreg$ to $\reg$. We can take $\zeta^\mu$ to be non-vanishing (corresponding to non-trivial $f$) only within a small region $|x_E^0| < \ell$ (for some $\ell \ll \tau$, but much larger than the cutoff). Of course, such a diffeomorphism has a non-trivial action on the background metric
\beq
g = (f^{-1})^*\eta
\eeq
where $*$ denotes the pullback. We claim that the reduced density matrix over $\dreg$ (with the metric $\eta$) is given by
\beq\label{dm}
\rho^{\psi}_{\dreg,\eta} = U^{\dagger}\,\rho^{\psi}_{\reg, g}\, U
\eeq
where $U$ is a unitary transformation, and $\rho^{\psi}_{\reg, g}$ is the reduced density matrix over the undeformed half-space, but with the deformed metric $g$.

We now give a quick formal proof of this claim. If we denote the eigenstates of $\phi$ restricted to $\dreg$ by $|\alpha\rangle, |\beta\rangle \cdots \in \mathfrak{h}_{\dreg}$, then we can construct a \emph{unitary}\footnote{The unitarity follows from the diffeomorphism invariance of the measure: $\left(f^*\right)^*\left[D\alpha\right]_{\eta} \equiv  \left[D \left(f^*\alpha_0\right)\right]_{\eta} = \left[D\alpha_0\right]_{g}.$ } operator $U: \mathfrak{h}_{\dreg} \to \mathfrak{h}_{\reg}$ given by
\beq
U  =\int [D\alpha]_{\eta}\, |( f^{-1})^*\alpha\rangle\langle \alpha|
\eeq
Then the claim \eqref{dm} can be checked explicitly by a series of manipulations on the path-integral definitions of the above density matrices \cite{Banerjee:2011mg}
\beqn\label{eq:diff_rho}
\langle\alpha| \rho^{\psi}_{\dreg,\eta} |\beta\rangle
&=& {\cal N}_{\dreg,\eta}^{-1}\int^{\phi^+=\beta}_{\phi^-=\alpha}[D\phi]_\eta\; X\;e^{-S\left[\eta, \phi\right]}\nonumber\\
&=&{\cal N}_{\dreg,\eta}^{-1} \int^{(f^*\tilde{\phi})^+=\beta}_{(f^*\tilde{\phi})^-=\alpha}[D (f^*\tilde{\phi})]_\eta\;X\;e^{-S\left[\eta, \left(f^*\tilde{\phi}\right)\right]}\nonumber\\ 
&=&{\cal N}_{\reg,g}^{-1} \int^{\tilde{\phi}^+=(f^{-1})^* \beta}_{\tilde{\phi}^-=(f^{-1})^*\alpha}[D\tilde{\phi}]_g \;X\;e^{-S\left[g, \tilde{\phi}\right]}\nonumber\\
&=&\langle (f^{-1})^*\alpha| \rho^{\psi}_{\reg,g} |(f^{-1})^*\beta\rangle\nonumber\\
&=&\langle \alpha| U^\dagger\rho^{\psi}_{\reg,g}U |\beta\rangle.
\eeqn
The first equality follows from the definition of $\rho_{A,\eta}^{\psi}$, the second equality is obtained by changing variables $\phi = f^*\tilde{\phi}$ inside the path integral, while the third equality follows from the assumption that the measure is diffeomorphism invariant. We have throughout used the fact that the operator insertions (denoted by $X$, following the definition \eqref{Oinsertions}) are away from the region where the diffeomorphism $f$ has non-trivial support, and so $f$ acts trivially on these operators. 

In the case where $f$ is an infinitesimal diffeomorphism, we can obtain a perturbative formula for $\rho^{\psi}_{\reg,g}$.  Writing the deformed metric on the Euclidean section as
\beq
g_{\mu\nu} = \delta_{\mu\nu}+2\pa_{(\mu}\zeta_{\nu)}+O(\zeta^2)
\eeq
where $\zeta$ is appropriately Wick rotated to Euclidean space, we obtain
\beq\label{ddm}
U\rho_{\dreg,\eta}^{\psi}U^{\dagger} = \rho_{\reg,\eta}^{\psi} + \frac{1}{2}\int d^dx\,\delta g_{\mu\nu}(x)\rho_{\reg,\eta}\left\{\frac{\mathcal{T}\left(T^{\mu\nu}(x)X\right)}{\left\langle X\right\rangle}-\frac{\left\langle T^{\mu\nu}(x)X\right\rangle X}{\left\langle X\right\rangle^2}\right\} +O(\zeta^2)
\eeq
where $\delta g_{\mu\nu}=2\pa_{(\mu}\zeta_{\nu)}$, and $\mathcal{T}$ is the angular-ordering operator: if $\theta \in (0,2\pi)$ is the angular coordinate in the $(x_E^0,x^1)$ plane, then
\beq
\mathcal{T}\left( \cO_a(\theta_a)\cO_b(\theta_b) \right)=  \cO_a(\theta_a)\cO_b(\theta_b)\,H(\theta_a-\theta_b) + \cO_b(\theta_b)\cO_a(\theta_a)\,H(\theta_b-\theta_a)
\eeq
where $H$ is the Heaviside step function. For the special case $|\psi\rangle = |0\rangle$, we then obtain 
\beq\label{vddm}
U\rho_{\dreg,\eta}U^{\dagger} = \rho_{\reg,\eta} + \frac{1}{2}\int d^dx\,\delta g_{\mu\nu}(x)\rho_{\reg,\eta}\Big(T^{\mu\nu}(x)-\left\langle T^{\mu\nu}(x)\right\rangle \Big) +O(\zeta^2)
\eeq

\subsection{Modular Hamiltonian}
\label{subsec:modham}
We are now in a position to construct the modular Hamiltonian over the deformed half-space for the vacuum state
\beq
K_{\dreg,\eta} \equiv -\mathrm{ln}\,\rho_{\dreg,\eta} = -U^{\dagger}\left(\ln\rho_{\reg,g}\right)U=U^{\dagger}K_{\reg,g}\,U
\eeq
In order to perturbatively expand the right hand side in powers of $\zeta$, we use the resolvent trick
 \beq
-\ln\rho_{\reg,g}=\int_0^\infty d\lambda\,\Big( \frac{1}{\rho_{\reg,g}+\lambda}-\frac{1}{1+\lambda}\Big)
\eeq
which together with equation \eqref{vddm} gives 
\beq
K_{\reg,g} = K_{\reg,\eta}+\delta_{\zeta}K_{\reg}+O(\zeta^2)
\eeq
\beq
\label{dln}
\delta_{\zeta}K_{\reg} = -\frac{1}{2}\int_0^{\infty}d\lambda\int d^{d}x\,\delta g_{\mu\nu}(x)\,\rho_{\reg,\eta}\,\frac{1}{\rho_{\reg,\eta}+\lambda}:T^{\mu\nu}:(x)\frac{1}{\rho_{\reg,\eta}+\lambda}
\eeq
where we have defined 
\beq
:T^{\mu\nu}: (x) = T^{\mu\nu}(x)-\langle T^{\mu\nu}(x)\rangle.
\eeq
In the interest of simplifying notation, we will henceforth drop the explicit reference to the Minkowski metric on $\rho_{\reg,\eta}$, and simply refer to it as $\rho_{\reg}$. It is possible to perform the $\lambda$ integral by going to the spectral representation (for details see \cite{Faulkner:2014jva, Faulkner:2015csl}, where similar calculations were performed). The result is
\beq
\label{dlnints}
\delta_{\zeta}K_{\reg}=\frac{1}{2} \int_{-\infty}^{\infty}ds  \;\frac{1}{4\sinh^2\left(\frac{s+i\epsilon}{2}\right)}
\int d^dx\,\delta g^{\mu\nu}(x)  \rho_{\reg}^{-is/2\pi}:T_{\mu\nu}:( x)\,\rho_{\reg}^{is/2\pi}
\eeq
Since the operator $\rho_{\reg}^{is/2\pi}$ generates modular evolution in Rindler time $s$, we see that the stress tensor is effectively liberated from the Euclidean section and inserted in real time. 

We now artificially split the integration region over which the stress tensor is inserted into two parts: a small solid cylinder $R_b$ of radius $b$ around the entangling surface, and its complement $\widetilde{R}$. We will
later show that the contribution from inside the cylindrical neighborhood vanishes in the limit $b \rightarrow 0$.
The region of integration is thus $R=   R_b \cup \widetilde{R}$ where we should remember that
$R$ contains a branch cut along the surface $A_0$.  We now write $\delta g_{\mu\nu} = \partial_\mu \zeta_\nu + \partial_\nu \zeta_\mu$ and integrate by parts
on the region $\widetilde{R}$
\beq\label{tospec}
\delta_{\zeta}K_{\reg} =   \int_{-\infty}^{\infty} d s \frac{1}{4\sinh^2\left(\frac{s+i\epsilon}{2}\right)}  \rho_{\reg}^{ -is/2\pi } \left(  - \int_{\widetilde{R}}  \left( \partial^\mu T_{\mu\nu} \right) \zeta^\nu  + \int_{\partial \widetilde{R}} :T_{\mu\nu}: \zeta^\nu d \Sigma^\mu \right) \rho_{\reg}^{ i s/2\pi} + \delta _\zeta K_b
\eeq
The first term involves the divergence of the stress tensor; in the absence of other operator insertions in the region where $\zeta$ has support, we can drop this term. (Indeed, expectation values in states of the form \eqref{states} which we will be interested in have precisely this property, since $\zeta$ has no support at the location of the operators $\mathcal{O}_{\alpha}$.)  The second term is integrated over $\pa \widetilde{R} = \pa R_b \cup \pa \widetilde{R}_+\cup \pa \widetilde{R}_-$ and gets two types of contributions: (i) from the boundary $\pa R_b$ of the hole of radius $b$, which we refer to as the \emph{imaginary cutoff surface}\footnote{Not to be confused with the UV cutoff surface that we
discuss in Appendix~\ref{app:cutoff}.}, and (ii) from $\pa \widetilde{R}_{\pm}$ above and below the region $\reg$, which we will refer to as the \emph{branch cut} (see figure \ref{fig:fig3noUVv2.pdf}). Finally $\delta_{\zeta} K_b$ represents
the contribution (iii) from inside the cylinder $R_b$.   

\myfig{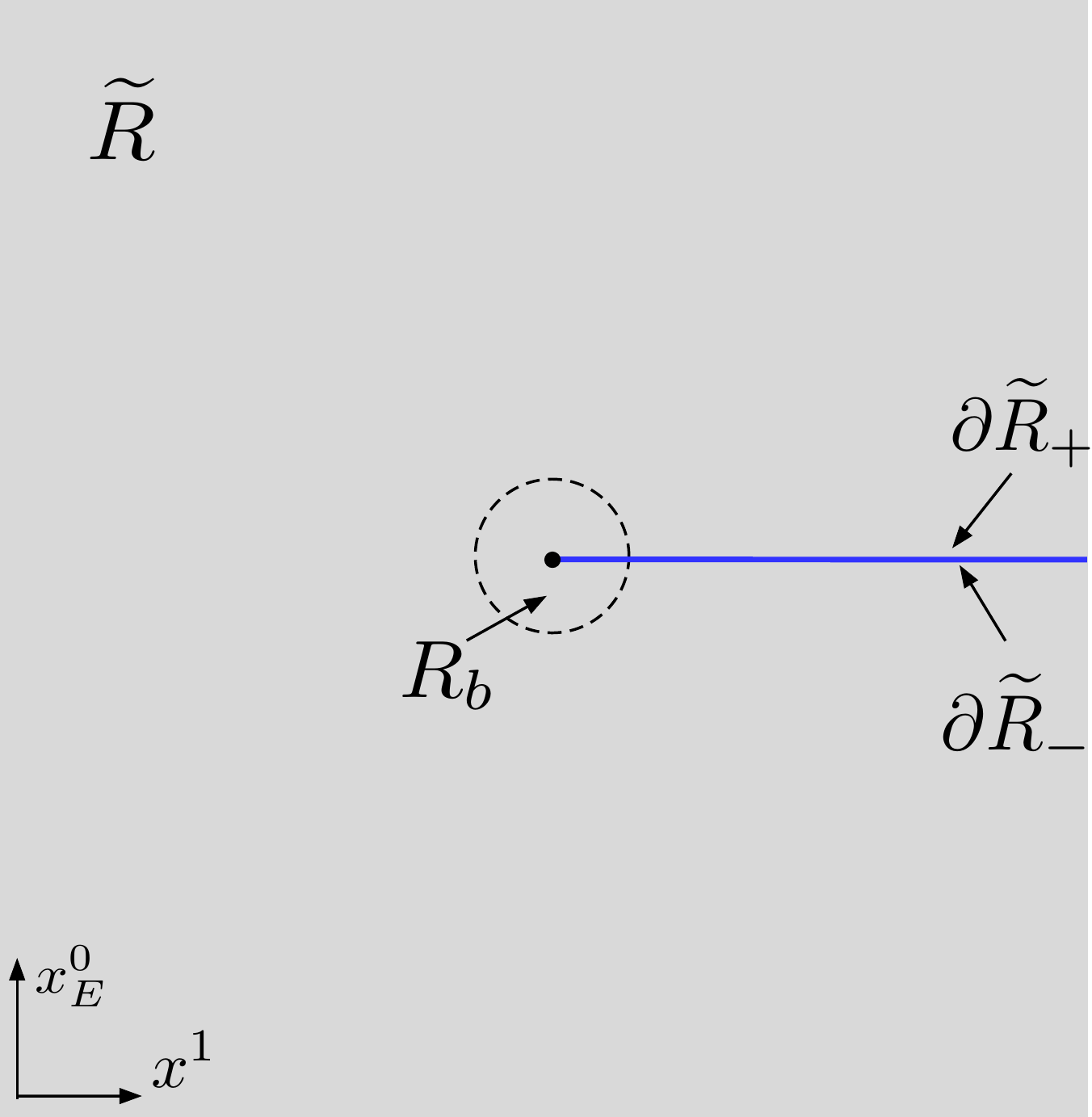}{6}{\small{\textsf{ We split the region of integration into two parts: the region inside the dotted line is $R_b$, and the region outside is $\widetilde{R}$. Also shown is the brach-cut $\pa \widetilde{R}_{\pm}$. }} }

(i) \emph{Imaginary cutoff surface}: Let us first deal with the term supported on the surface $\pa R_b$. It is convenient to switch to complex coordinates 
\beq
z= x^1-ix_E^0,\;\;\bar{z} = -(x^1+ix_E^0).
\eeq
In these coordinates, we find
\begin{align}
\label{tpp}
\rho_{\reg}^{-is/2\pi}\,\Big(\zeta^{\mu} n^{\nu}T_{\mu \nu}(x)\Big)\,\rho_{\reg}^{is/2\pi}  =   & \Big(-e^{2s-i\theta} T_{zz}(x_s)+T_{z\bar{z}}(x_s)e^{i\theta}\Big) \zeta^z  \\ & \qquad + \Big(-T_{z\bar{z}}(x_s)e^{-i\theta}+e^{-2s+i\theta}T_{\bar{z}\bar{z}}(x_s)\Big) \zeta^{\bar{z}}  \nonumber
\end{align}
where, 
\beq
x^{\mu}_s = \left(b \sin(\theta +is),\,b \cos(\theta +is),\,\vec{x}\right)
\eeq
Further, $n^\nu$ is the (inward pointing) unit normal to $\pa R_b$ 
\beq
n =  e^{i \theta} \partial_{\bar z} - e^{ - i \theta} \partial_z.
\eeq
and $\zeta^z$ and $\zeta^{\bar{z}}$ are the components of the vector field $\zeta$ in holomorphic coordinates
\beq
\zeta = \zeta^z \partial_z + \zeta^{\bar{z}} \partial_{\bar z}
\eeq
We now proceed by shifting the $s$ integration contour $s \rightarrow s + i \theta$ in order to remove the $\theta$ dependence from the stress tensor. We do this after switching the order
of integration so that the $s$ integral comes before the $\theta$ integral. This step assumes  analyticity in the complex $s$ plane and that the contributions from $s \rightarrow \pm \infty$ vanish, which can be justified in a spectral representation of \eqref{tospec}. This gives
\begin{align}
\left.\delta_{\zeta}K_{\reg}\right|_{\pa R_b}=-b \int d^{d-2}\vec{x}\int_0^{2\pi}d\theta \int_{-\infty}^{\infty} d s \frac{1}{4\sinh^2\left(\frac{s+i\theta}{2}\right)}  & \Big( (e^{2s}T_{zz}-:T_{z\bar{z}}:) \zeta^z e^{ i \theta} \Big. \\ & \qquad \Big. +  (:T_{z\bar{z}}: - e^{-2s}T_{\bar{z}\bar{z}}) \zeta^{\bar{z}} e^{ -i \theta} \Big)
\end{align}
where now these stress-tensors are evaluated at $x^0_E= i b \sinh(s), x^1 = b \cosh(s)$. We can now perform the $\theta$ integral using
\beq
\label{sint}
 \int_0^{2\pi} d \theta  \frac{1}{4\sinh^2\left(\frac{s+i\theta}{2}\right)} e^{ \pm i \theta}  = 2\pi e^{\mp s}\Theta(\pm s)-2\pi \delta (s)
\eeq
The delta function term above can be dropped since this term does not contribute in the limit $b\rightarrow 0$.\footnote{Actually, rather than drop this term, let us add it to a stack: 
$$\mathrm{Stack}=  2\pi b \int  d^{d-2} \vec{x}\,  T_{1\mu}\,\zeta^\mu. $$ We will update Stack everytime we find a term of this type in our calculation. \label{fn:Stack}} So we get
\beq
\label{prelim}
\left.\delta_{\zeta}K_{\reg}\right|_{\pa R_b}=2\pi b  \int d^{d-2}\vec{x}  \Big( -\int_0^\infty d s ( e^s T_{zz} - e^{-s} :T_{z\bar{z}}: ) \zeta^z 
+  \int_{-\infty}^0 d s ( e^{-s} T_{\bar{z}\bar{z}} - e^{s} :T_{z\bar{z}}:) \zeta^{\bar{z}}  \Big)
\eeq
Naively, it might seem that all the terms on the right hand side vanish in the $b \to 0$ limit. In fact, the terms involving $T_{z\bar{z}}$ do indeed vanish in this limit.\footnote{By this we mean that the $T_{z\bar{z}}$ terms do not contribute to matrix elements in the class of states \eqref{states} which are of interest here. On the other hand, if we were to evaluate matrix elements in Rindler eigenstates we would find potential divergences in this limit. Note that since we used a spectral representation for $\rho_{A_0}$ at an intermediate stage, we were exactly evaluating this in  Rindler eigenstates. So the order in which this limit is taken is a somewhat delicate issue which is best ignored on a first pass. In Appendix~\ref{app:cutoff} we confront this issue explicitly. } However, the terms involving $T_{zz}$ and $T_{\bar{z}\bar{z}}$ get an enhancement from the $s$ integral, coming from the $s \sim -\ln b$ and $s\sim \ln b$ limits respectively.
Taking the limit $b \rightarrow 0$ and Wick rotating the vector field back to real time, i.e. $\zeta^z \to \zeta^+$ and 
$\zeta^{\bar{z}} \to \zeta^-$ (where the light-cone coordinates are defined as $x^{\pm} = x^0\pm x^1$), we obtain
\beq
\label{exactly} 
\left.\delta_{\zeta}K_{\reg}\right|_{\pa R_b}=2\pi \int d^{d-2}\vec{x}\left( - \int_0^\infty d x^+ \zeta^+ T_{++}(x^+,x^-=0,\vec{x}) +\,\int_{-\infty}^0 d x^- \zeta^- T_{--}(x^+=0,x^-,\vec{x}) \right)
\eeq 
where note that the first term on the right hand side is integrated over the future Rindler horizon $\mathcal{H}_+$, while the second term is integrated over the past Rindler horizon $\mathcal{H}_-$, shown in figure \ref{fig:fig1.pdf}.

(ii) \emph{Branch cut}: Now we come to the second remaining term supported over $\pa \widetilde{R}_+\cup \pa \widetilde{R}_-$. Once again, deforming the $s$ contours to get rid of the $\theta$ dependence from the stress tensors, we obtain
\beq\label{res}
\left.\delta_{\zeta}K_{\reg}\right|_{\pa \widetilde{R}_{\pm}}= \int d^{d-2}\vec{x}\int_b^{\infty}dx^1\int_{-\infty}^{\infty} d s \left(-\frac{1}{4\sinh^2\left(\frac{s+i\epsilon}{2}\right)}+\frac{1}{4\sinh^2\left(\frac{s-i\epsilon}{2}\right)}\right) t^{\mu}\zeta^\nu \rho_{\reg}^{ -is } :T_{\mu\nu}: \rho_{\reg}^{ i s}
\eeq
where $t = \pa_{x^0_E}$, and the stress tensor is evaluated on the region $\reg$, i.e. $T_{\mu\nu} \equiv T_{\mu\nu}(x_E^0=0,x^1,\vec{x})$ above. The first term inside the brackets comes from $\pa \widetilde{R}_+$ while the second term comes from $\pa \widetilde{R}_-$ (after the contour deformation $s \to s +2\pi -2\epsilon$). 
It is clear from equation \eqref{res} that the $s$ integral precisely picks out the double-pole at $s=0$. A straightforward application of the residue theorem gives 
\beq
\label{comm}
\left.\delta_{\zeta}K_{\reg}\right|_{\pa \widetilde{R}_{\pm}} = 2\pi\int d^{d-2} \vec{x} \int_b^\infty dx^1\, t^{\mu}\zeta^\nu   \left[ T_{\mu\nu}(0,x^1,\vec{x}), K_{\reg}\right]
\eeq

(iii) \emph{Inside the hole}: We can follow the same methods as in (i). Pick coordinates
close to the entangling surface such that:
\beq
ds^2 = dr^2 + r^2 d\theta^2 + d\vec{x}^2 \rightarrow dr^2 - r^2 d s^2 + d\vec{x}^2
\eeq
where we have also shown the Wick rotated Rindler coordinates.  The hole region $R_b$ corresponds to $r<b$ and since we are again working in a region close to the entangling surface we can take the diffeomorphism at leading order to be independent of $r$. After shifting the integration contour $s \rightarrow s+ i \theta$ and Wick rotating the $\zeta$ vector field we have:
\beq
\label{insideb}
\rho_{A_0}^{ \frac{- i s+\theta}{2\pi} } T_{\mu\nu}(x) \partial^\mu \zeta^\nu 
\rho_{A_0}^{ \frac{i s- \theta}{2\pi}} = e^{ i \theta+s} T_{i +}(x^+,x^-, \vec{x}) \partial_i \zeta^+ 
+ e^{ - i \theta-s} T_{i -}(x^+,x^-, \vec{x}) \partial_i \zeta^-
\eeq
The light-like coordinates where the stress tensor on the right hand side above is located are $x^\pm =\pm r e^{\pm s}$. We still have to integrate \eqref{insideb} over:
\beq
\int d^d x \int ds \frac{1}{4 \sinh^2\left( \frac{s + i \theta}{2} \right)} \ldots = \int d^{d-2} \vec{x}  \int_{r<b} d r  r \oint d\theta \int_{-\infty}^\infty ds  \frac{1}{4 \sinh^2\left( \frac{s + i \theta}{2} \right)}  \ldots
\eeq
But at this point the $\theta$ dependence is
the same as in \eqref{sint} and we can again do the $\theta$ integral. After ignoring the $\delta(s)$ contribution which vanishes in the limit $b \rightarrow 0$\footnote{Once again, we add this term to the stack defined in footnote \ref{fn:Stack}:
$$\mathrm{Stack} \to  \mathrm{Stack} -2\pi \int d^{d-2} \vec{x} \int_0^b dx^1 x^1 T_{i \mu} \partial^i \zeta^\mu .$$ },  this has exactly the effect of switching
the angular integral in the Euclidean calculation to a real time integral localized near the Rindler horizon: $0< r < b$
and $ -\infty < s < \infty$ (see figure \ref{fig:fig4.pdf}). The integrand
is the stress tensor coupled to a real time diffeomorphism of the metric
\beq
\label{realb}
\delta_\zeta K_{b} = 2\pi \int_{0<r<b} d^d x\, T_{\mu\nu}\, \partial^\mu \widetilde{\zeta}^\nu 
\eeq
for the following vector field:
\beq
\label{realb2}
\widetilde{\zeta} = \Theta(x^0)  \zeta^+ \partial_+ + \Theta( - x^0) \zeta^- \partial_-
\eeq
We have again ignored a contribution localized at $x^0=0$, coming from the derivative
of the step functions above,  which vanishes in the limit $b \rightarrow 0$.\footnote{These terms go into the stack as well:$$\mathrm{Stack} \to  \mathrm{Stack} + 
2\pi \int d^{d-2}\vec{x} \int_0^b d x^1 ( T_{10} \zeta^0 + T_{00} \zeta^1 )
= 2\pi \int d^{d-2} \vec{x} \int_0^b d x^1 \left( x^1 \partial_0 T_{0\mu}\zeta^\mu + 2 T_{10} \zeta^0 
+ (T_{11} +T_{00}) \zeta^1 \right)$$
where in the second equality we have integrated by parts; this is then exactly the extension of the $x^1$ integral in \eqref{comm} so that
it ranges from $0$ to $\infty $. Even though all these terms vanish as $b \rightarrow 0$, it is satisfying that they add up like this. }

\myfig{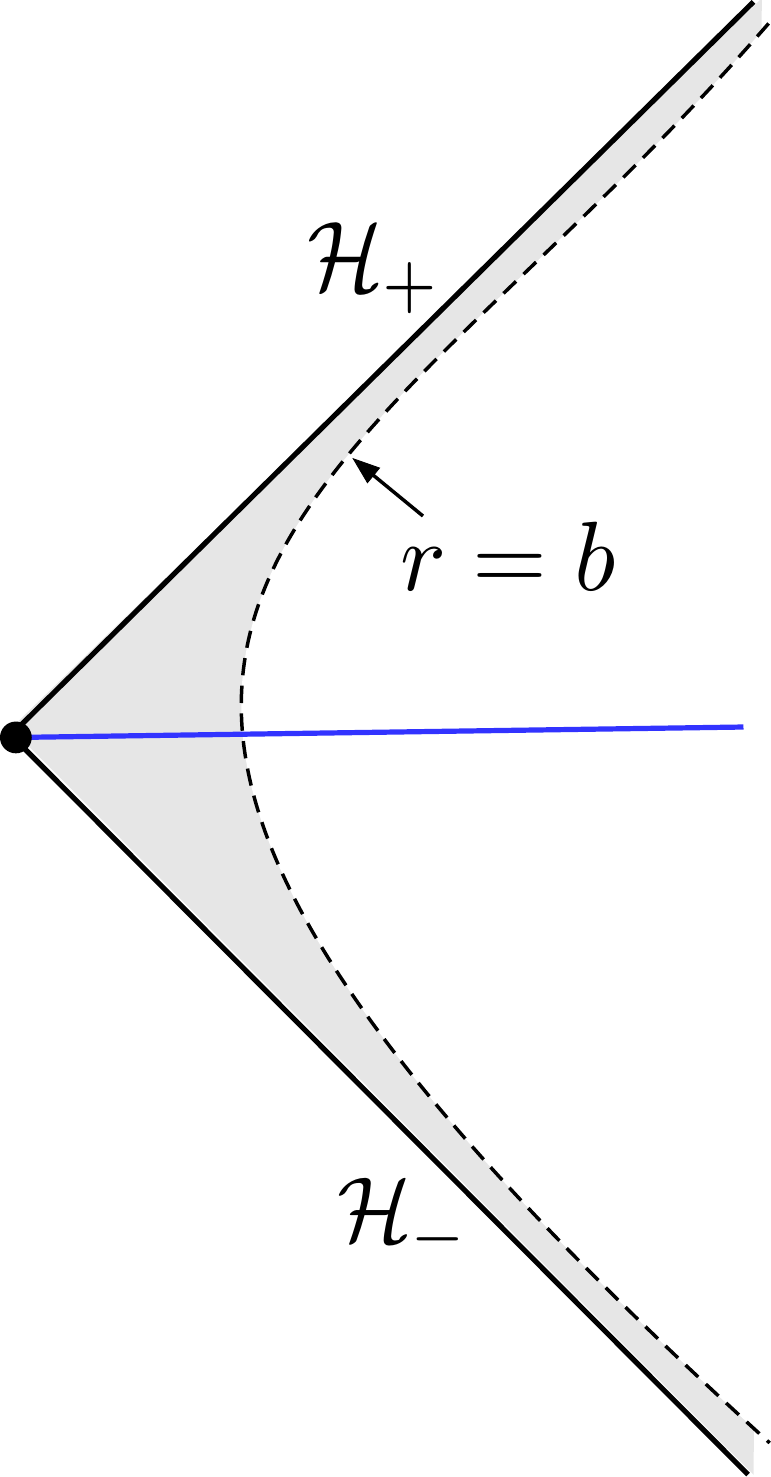}{3.5}{\small{\textsf{ The contribution from inside the region $R_b$ can be written in real time as an integral over the shaded region.  }} }

It is not hard to see that \eqref{realb} should vanish in the limit $b \rightarrow 0$. However it
is somewhat enlightening to go another route and instead integrate by parts on \eqref{realb}. 
We get two terms, one from the $r=b$ boundary and the other from precisely the past and future Rindler horizons on the boundary of the domain of dependence of $A_0$.
It turns out the former term cancels \eqref{prelim} prior to taking the $b \rightarrow 0$ limit (although we always need $b$ small), and the later term is \emph{exactly} the desired result given in \eqref{exactly} . So in the end when we add all the terms together, no $b \rightarrow 0$ limit is necessary and 
the null energy operators in  \eqref{exactly} simply emerge.  This is perhaps not too surprising since the $r=b$ surface is imaginary, and there should be no dependence on $b$, however we find the detailed cancelations that occur and the form in \eqref{realb} intriguing (including in the running footnote Stack), perhaps hinting that there is a different way to do this calculation directly in real times.

To summarize, putting everything together, we find that the modular Hamiltonian over the deformed half-space is given by
\beqn
UK_{\dreg}U^{\dagger}
&=& K_{\reg}-2\pi \int_{\mathcal{H}_+} \zeta^+ T_{++}+ 2\pi \int_{\mathcal{H}_-} \zeta^- T_{--} +2\pi \int_{\reg}  t^{\mu}\zeta^\nu\left[ T_{\mu\nu}, K_{\reg} \right]\label{mH1}
\eeqn
which is the result claimed in \eqref{mH}.\footnote{Roughly speaking, the ``null-energy'' terms measure the amount of modular energy leaving the Rindler wedge, while the commutator term comes from the action of the unitary transformations on the original (undeformed) modular Hamiltonian.} We can also derive a similar expression for the modular Hamiltonian corresponding to the complement $\dregc$
\beqn
VK_{\dregc}V^{\dagger}
&=& K_{\regc}+2\pi \int_{\mathcal{H}^c_+} \zeta^+ T_{++}- 2\pi \int_{\mathcal{H}^c_-} \zeta^- T_{--} +2\pi \int_{\reg^c}  t^{\mu}\zeta^\nu\left[ T_{\mu\nu},K_{\regc} \right]\label{mH2}
\eeqn
where $\mathcal{H}_{\pm}^c$ are the Rindler horizons corresponding to the complement $\regc$, and $V: \mathfrak{h}_{\dregc}\to \mathfrak{h}_{\regc}$ is a unitary transformation. Finally, putting these together, we obtain the following formula for the full modular Hamiltonian
\beqn\label{fmH}
\mathcal{U}\cK_{\dreg}\mathcal{U}^{\dagger}
&=& \cK_{\reg}-2\pi \int_{\mathcal{L}_+} \zeta^+ T_{++}+ 2\pi \int_{\mathcal{L}_-} \zeta^- T_{--} +2\pi \int_{\Sigma}  t^{\mu}\zeta^\nu\left[T_{\mu\nu}, \cK_{\reg} \right]
\eeqn
where we have defined the light sheets $\cL_{\pm} = \mathcal{H}_{\pm}\cup\mathcal{H}_{\pm}^c$, and $\mathcal{U}:\mathfrak{h}_{\Sigma}\to \mathfrak{h}_{\Sigma}$ is a unitary transformation given by $\mathcal{U} = U\otimes V$. 

\section{Averaged Null Energy Condition}

In this section, we will consider the expectation value $\langle \psi |\cK_{\dreg}|\psi\rangle $ in the class of states \eqref{states}. We will then use the positivity of the operator $\cK_{\reg}-\cK_{\dreg}$ to prove the averaged null energy condition within this class. 

\subsection{Positivity of $\cK_{\reg}-\cK_{\dreg}$}
For completeness, we begin with a brief review of the argument that $\cK_\reg - \cK_\dreg$ is a positive operator, following \cite{Blanco:2013lea}.\footnote{Similar arguments have been used in
\cite{Wall:2009wi,Bousso:2014uxa}. A rigorous proof of the positivity of this operator can also be found in \cite{zbMATH00845667} which uses methods of algebraic QFT. } Consider any two states, which we take here to be the vacuum $|0\rangle$ and a non-trivial pure state $|\psi\rangle$. Given an entangling region $\reg$ and the corresponding reduced density matrices $\rho_{\reg}$ and $\rho^{\psi}_{\reg}$, one defines the relative entropy
\beqn
S(\rho^{\psi}_{\reg} || \rho_{\reg}) 
&=& \mathrm{Tr}_{A_0}\left(\rho^{\psi}_{\reg}\,\mathrm{ln}\,\rho^{\psi}_{\reg}\right)-\mathrm{Tr}_{A_0}\left(\rho^{\psi}_{\reg}\,\mathrm{ln}\,\rho_{\reg}\right)\\
&=& \left[\mathrm{Tr}_{A_0}\left(\rho^{\psi}_{\reg}\,K_\reg \right)-\mathrm{Tr}_{A_0}\left(\rho_{\reg}\,K_\reg\right)\right]+
 \left[\mathrm{Tr}_{A_0}\left(\rho^{\psi}_{\reg}\,\mathrm{ln}\,\rho^{\psi}_{\reg}\right)-\mathrm{Tr}_{A_0}\left(\rho_{\reg}\,\mathrm{ln}\,\rho_{\reg}\right)\right]\nonumber\\
 &\equiv& \Delta \langle K_\reg\rangle-\Delta S_{EE}[\reg].\label{RE}
\eeqn
where $K_\reg$ is the modular Hamiltonian corresponding to the vacuum state over the region $\reg$.

Relative entropy has a number of interesting properties. For instance, it is a positive quantity
\beq
S(\rho^{\psi}_{\reg} || \rho_{\reg})  \geq 0
\eeq
Further, if we pick another region $\dreg$ such that $\dreg \subset \reg$ (more precisely, if $\mathcal{D}(\dreg) \subset \mathcal{D}(\reg)$, where $\mathcal{D}(A)$ is the domain of dependence of $A$) then the monotonicity of relative entropy implies
\beq\label{mon}
S(\rho^{\psi}_{\dreg} || \rho_{\dreg}) \leq S(\rho^{\psi}_{\reg} || \rho_{\reg})
\eeq
Intuitively, the relative entropy measures the distinguishability between two states. From this point of view, the monotonicity property states that the distinguishability between two states decreases as we consider their reduced density matrices over smaller and smaller regions.\footnote{See \cite{Lashkari:2014yva, Lashkari:2015dia, Sarosi:2016oks} for field theoretic calculations of relative entropy in excited states, using a version of the replica trick.}

From equations \eqref{RE} and \eqref{mon}, we obtain
\begin{align}
\Delta \left< K_\dreg \right> -\Delta \left< K_\reg \right>  - \Delta S_{EE}[\dreg] +  \Delta S_{EE}[\reg]  \leq 0 \\
\Delta \left< K_{\regc} \right> -\Delta \left< K_{\dregc} \right>  - \Delta S_{EE}[\regc] +  \Delta S_{EE}[\dregc]  \leq 0
\end{align}
where all modular Hamiltonians are defined relative to the vacuum. Adding the two inequalities we have
\beq
\Delta \left< \cK_\dreg \right> -\Delta \left< \cK_\reg \right> - \Delta S_{EE}[\dreg] +  \Delta S_{EE}[\reg]   - \Delta S_{EE}[\regc] +  \Delta S_{EE}[\dregc]   \leq 0 
\eeq
Now, since all vacuum contributions vanish, we can drop the $\Delta$. (This is because $\cK_A$ annihilates the vacuum for any region $A$). This implies
\beq
\left< \cK _\dreg \right>_\psi - \left< \cK_\reg \right>_\psi \leq S_{EE}[\psi,\dreg] - S_{EE}[\psi,\dregc] - S_{EE}[\psi,\reg]+ S_{EE}[\psi,\regc] = 0
\eeq
where the last equality follows from the purity of $|\psi\rangle$. Since this is true for any pure state $|\psi\rangle$, we deduce that $\cK _\reg - \cK _\dreg$ is a positive operator.

\subsection{Computing $\langle \cK_{\reg}\rangle_{\psi}-\langle \cK_{\dreg}\rangle_{\psi}$}
\label{subsec:comp}
We now explicitly compute the expectation value of $\cK_{\reg}-\cK_{\dreg}$ in the state
\beq\label{sts}
|\psi\rangle =  e^{-\tau H}\sum_{\a} c_{\a}  \mathcal{O}_{\a}(0,\bx)|0\rangle
\eeq
where as before we take $\reg$ to be the half-space $x^1>0$, and $\dreg$ to be the deformed half-space. Using the relation
\beq
\langle \psi|\cK_{\dreg}|\psi \rangle = \mathrm{Tr}_{\dreg}\left(\rho_{\dreg}^{\psi}K_{\dreg}\right)-\mathrm{Tr}_{\dreg^c}\left(\rho_{\dreg^c}^{\psi}K_{\dreg^c}\right)
\eeq
we find
\beq
\langle \psi|(\cK_{\reg}-\cK_{\dreg})|\psi \rangle =\mathfrak{T}^{(1)}+\mathfrak{T}^{(2)}
\eeq
where
\beq\label{t1}
\mathfrak{T}^{(1)}=- \mathrm{Tr}_{\reg}\left(\delta_{\zeta}\rho_{\reg}^{\psi}K_{\reg}\right)+\mathrm{Tr}_{\reg^c}\left(\delta_{\zeta}\rho_{\reg^c}^{\psi}K_{\regc}\right)
\eeq
\beq\label{t2}
\mathfrak{T}^{(2)}=-\mathrm{Tr}_{\reg}\left(\rho_{\reg}^{\psi}\delta_{\zeta}K_{\reg}\right)+\mathrm{Tr}_{\reg^c}\left(\rho_{\reg^c}^{\psi}\delta_{\zeta}K_{\regc}\right)
\eeq
where note that the unitary transformations $U$ and $V$ have dropped out inside the trace.
The second term above is straightforward to evaluate from equations \eqref{mH1} and \eqref{mH2}:
\beq\label{t2'}
\mathfrak{T}^{(2)} = 2\pi \int_{\mathcal{L}_+} \zeta^+ \langle T_{++}\rangle_{\psi}- 2\pi \int_{\mathcal{L}_-} \zeta^- \langle T_{--}\rangle_{\psi} + 2\pi \int_{\Sigma}  t^{\mu}\zeta^\nu\langle\left[ \cK_{\reg}, T_{\mu\nu} \right]\rangle_{\psi}
\eeq
where we have defined
\beq
\langle Y \rangle_{\psi}\equiv\frac{\left\langle \psi |Y|\psi \right\rangle}{\langle \psi | \psi\rangle}.
\eeq
The first two terms in \eqref{t2'} are the appropriate null energy expectation values (albeit integrated along the transverse $\vec{x}$ directions with arbitrary coefficients $\zeta^{\pm}(\vec{x})$) which enter in the averaged null energy condition. We will see below that the last term in \eqref{t2'} is precisely cancelled by a contribution coming from $\mathfrak{T}^{(1)}$. 

Consider for instance, the first term in  $\mathfrak{T}^{(1)}$; from equations \eqref{ddm} and \eqref{t1} we obtain
\beq
\mathrm{Tr}_{\reg}\left(\delta_{\zeta}\rho_{\reg}^{\psi}K_{\reg}\right) =  \frac{1}{2}\int_{R} \delta g_{\mu\nu}(x)\Big\{\frac{\left\langle T^{\mu\nu}(x)K_\reg X\right\rangle}{\langle X\rangle}-\frac{\langle T^{\mu\nu}(x)X\rangle\langle  K_\reg X\rangle}{\langle X\rangle^2}\Big\}
\eeq
On the right hand side, the correlators $\langle \cdots\rangle $ indicate Euclidean correlation functions on $\re^d$, and recall the notation
\beq
X=\sum_{\a}\sum_{\a'}c^*_{\a}c_{\a'} \cO_{\a}^{\dagger}(\tau,\bx) \cO_{\a'}(-\tau,\bx).
\eeq
Note that the Euclidean correlation function appears naturally from the path integral
construction of the deformed density matrix for the excited state -- we need only
add an insertion of $K_{A_0}$ along $A_0$ and trace -- resulting in the above correlation function. 
Because of the $K_{A_0}$ operator insertion, we should remove an infinitesimal cut running along $A_0$  from the region over which we integrate the diffeomorphism $R$.   Bearing this in mind, we can integrate by parts in the region $R$ to obtain only a contribution from above and below
the $K_{A_0}$ operator insertion, yielding a commutator:
\beq
\mathrm{Tr}_{\reg}\left(\delta_{\zeta}\rho_{\reg}^{\psi}K_{\reg}\right) =  
 -2\pi \int_{\Sigma}d^{d-2}\vec{x}\int_{0}^{\infty}dx^1 \,\zeta_{\mu}t_{\nu}\left\langle [K_\reg,T^{\mu\nu}(0,x^1,\vec{x})] \right\rangle_{\psi}
\eeq
where we reiterate that $\zeta$ vanishes at the locations of the operators $\cO_{\alpha}$, which allows us to drop the divergence of the stress tensor. For the full modular Hamiltonian we then have:
\beq
\mathfrak{T}^{(1)}=  -2\pi \int_{\Sigma}d^{d-2}\vec{x}\int_{-\infty}^{\infty}dx^1 \,\zeta_{\mu}t_{\nu}\left\langle [\cK_\reg,T^{\mu\nu}(0,x^1,\vec{x})] \right\rangle_{\psi}
\eeq
As promised, this term precisely cancels the last term in equation \eqref{t2'}. We therefore conclude that
\beq\label{eq:full_modular}
\langle \cK_{\reg}\rangle_{\psi} - \langle \cK_{\dreg}\rangle_{\psi} = 2\pi \int_{\mathcal{L}_+} \zeta^+ \langle T_{++}\rangle_{\psi}-2\pi \int_{\mathcal{L}_-} \zeta^- \langle T_{--}\rangle_{\psi} 
\eeq
Since $\zeta^+>0$ and $\zeta^-<0$ by construction (see figure \ref{fig:fig1.pdf}), the positivity of the operator $(\cK_{\reg}-\cK_{\dreg})$ leads to the averaged null energy conditions
\beqn
 \int_{-\infty}^{\infty}dx^+  \left\langle T_{++}(x^+,x^-=0,\vec{x})\right\rangle_{\psi} & \geq & 0,\nonumber\\
  \int_{-\infty}^{\infty}dx^-  \left\langle T_{--}(x^+=0,x^-,\vec{x})\right\rangle_{\psi}  &\geq & 0
\eeqn
This concludes our proof that the monotonicity property of $\cK_{\dreg}$ implies the ANEC. 
While we presented the proof above in the context of half-spaces in Minkowski space-time, the above calculation can also be extended in an obvious way to general \emph{static} bifurcate Killing horizons. In this case we would be studying the modular Hamiltonian for small deformations of the entangling cut away from the bifurcation point in the Hartle-Hawking state.  The monotonicity constraint then leads to the ANEC for complete null generators of the Killing horizon.

\section{Modular Hamiltonians in AdS/CFT}
\label{sec:jlms}
In this section we make a connection between our results and the recently proposed JLMS formula \cite{Jafferis:2015del}:
\beq\label{eq:JLMS}
K_{A}^{\text{CFT}}=\frac{\mathrm{Area}_{\partial \mathcal{M}\backslash A}}{4 G_N}+K_{\mathcal{M}}^{ \text{bulk}}+\cdots+O(G_N) 
\eeq
where $A$ denotes the boundary subsystem, and $\mathcal{M}$ denotes the bulk region enclosed by the minimal/extremal Ryu-Takayanagi/HRT \cite{Ryu:2006bv,Hubeny:2007xt} surface $\pa \mathcal{M}\backslash A$ (which ends on the entangling surface $\pa A$, and is homologous to $A$). Further, the $\cdots$ denote local terms on the extremal surface, which will not be relevant in the following discussion. The result \eqref{eq:JLMS}
arises as a consequence of the formula for quantum corrections to the Ryu-Takaynagi entropy \cite{Faulkner:2013ana,engelhardt2015quantum} (see also \cite{Dong:2016eik}.)

Here, we want to perform a simple consistency check of our formula \eqref{fmH} for the full modular Hamiltonian of deformed half-spaces against equation (\ref{eq:JLMS}).  
In particular, restricting to pure states so that the bulk area operator $\text{Area}_{\pa \mathcal{M}\backslash A}$ evaluates the same over $A$ and $A^c$, (\ref{eq:JLMS}) allows us to equate the full modular energies between the boundary and the bulk theories
\beq \label{jlms2}
\cK_{A}^\text{CFT}=\cK_{\mathcal{M}}^\text{bulk}+\mathcal{O}(G_N)
\eeq
where note that the local terms on the extremal surface have also dropped out. 

In order to make contact with our previous results, we take $A$ to be a small deformation of the boundary half-space $\reg = \{x^{1}>0,\,x^0=0\}$. If we use the coordinates $\left(z,x^0, x^1,\vec{x}\right)$ on the Poincar\'e patch of AdS, with 
\beq
g_{AdS} = \frac{dz^2-(dx^0)^2+(dx^1)^2+(d\vec{x})^2}{z^2},
\eeq
then the corresponding (undeformed) extremal surface in the bulk is given by the codimension two surface $x^0=x^1=0$, and we have the corresponding undeformed region $\mathcal{M}_0= \{x^{1}>0, x^0=0\}$. To linear order in the CFT shape deformation $\zeta$, we then expect
\beq \label{jlms3}
\langle \delta_{\zeta} \cK_{\reg}^\text{CFT} \rangle_{\psi_\text{CFT}}=\langle\delta_{\zeta_\text{bulk}} \cK_{\mathcal{M}_0}^\text{bulk}\rangle_{\psi_\text{bulk}}+\mathcal{O}(G_N) ,
\eeq   
where we have evaluated the deformations in the CFT and bulk modular Hamiltonians in an excited boundary state $|\psi\rangle_\text{CFT}$ and the dual bulk state $|\psi\rangle_\text{bulk}$ respectively. Further, $\zeta_\text{bulk}$ is the deformation in the bulk minimal surface as a consequence of the boundary shape deformation, and approaches $\zeta$ in the limit $z\to 0$; it is fixed away from the boundary by the requirement that the deformed bulk surface remain extremal \cite{Hubeny:2007xt}.  

For simplicity, we focus on null deformations on the CFT side, i.e. $\zeta^-=0$. The left-hand side of equation \eqref{jlms3} has been computed previously; see equation \eqref{eq:full_modular}. Furthermore, if we view the bulk effective theory as a weakly coupled quantum field theory in background AdS geometry, then we expect that our flat space arguments from the previous sections can be extended to the bulk effective theory -- this is because we have a Killing vector field in AdS which generates a boost around the unperturbed extremal surface in the $(x^0, x^{1})$ plane. The bulk modular Hamiltonian then satisfies a covariant version of (\ref{eq:full_modular}):
\beq\label{eq: bulk_full_mod}
\langle \delta_{\zeta_\text{bulk}}\cK_{\mathcal{M}}^\text{bulk}\rangle_{\psi_\text{bulk}}= -\int_{\mathcal{L}^+(\partial{\mathcal{M}_0})} \sqrt{h} \;\zeta^+_\text{bulk}(z,\vec{x})\langle T^\text{bulk}_{++}(x^+,z,\vec{x})\rangle_{\psi_\text{bulk}} \nonumber\\,
\eeq
where $h_{ij}$ is the induced metric on the undeformed minimal surface $\partial{\mathcal{M}_0}$:
\beq
h = \frac{dz^2+(d\vec{x})^2}{z^2}.
\eeq
A consequence of JLMS combined with our calculations is therefore:
\beq\label{eq: full_mod_equal}
\int_{\mathcal{L}^+(\partial{\reg})} d^{d-2}\vec{x} dx^+\; \zeta^+\langle \psi_\text{CFT}| T^\text{CFT}_{++} |\psi_\text{CFT}\rangle =\int_{\mathcal{L}^+(\partial{\mathcal{M}_0})} \sqrt{ h}  d^{d-2}\vec{x} dx^+dz\; \zeta^+_\text{bulk}\langle \psi_\text{bulk}| T^\text{bulk}_{++}|\psi_\text{bulk}\rangle
\eeq
Our goal now is to establish this equivalence using the usual rules of AdS/CFT \cite{maldacena1999large, Witten:1998qj,Gubser:1998bc}.
In particular in order to apply the JMLS argument the state under consideration must
have a  perturbative in $G_N$ back-reaction on the bulk $AdS$ space-time and so we will
consider boundary states of the form $|\psi_\text{CFT}\rangle=O(x_\psi)|0\rangle_\text{CFT}$, where $O$ is a single trace primary scalar operator of dimension $\Delta = \mathcal{O}(1)$ in terms of large-$N$
counting. The operator is inserted at $x_\psi~:~(x^0_\psi = -i \tau, x^1_\psi =0,  \vec{x}_\psi = 0)$.  In this case the dual bulk state can be identified as $|\psi_\text{bulk}\rangle=\lim_{z\to 0} z^{-\Delta}\phi(z,x_\psi)|0\rangle_\text{AdS}$ for the corresponding bulk field $\phi$. The equality in \eqref{eq: full_mod_equal} can already be read off from the work of Hofman and Maldacena \cite{Hofman:2008ar} but here, for completeness, we will give another derivation. 

For a CFT with a weakly coupled Einstein gravity dual in the bulk, the LHS can be computed using Witten diagrams. In particular, we assume that the relevant part of the bulk Lagrangian takes the form: 
\beq
\mathcal{S}^{\text{bulk}} \sim \int_{\text{AdS}_{d+1}}\sqrt{G} \left[-\frac{1}{G_N}(R_G+\Lambda) + \frac{1}{2}G^{\mu\nu}\partial_\mu \phi \partial_\nu \phi - m_\Delta^2\phi^2\right],\;m_\Delta^2=\Delta(\Delta-d)
\eeq  
The leading order Witten diagram contribution to the LHS of (\ref{eq: full_mod_equal}) comes from the coupling between the graviton ($h^{\mu\nu}=\delta G^{\mu\nu}$) and the scalar stress tensor: 
\beq
\mathcal{L}^\text{bulk}_\text{int}=\sqrt{G} h^{\mu\nu}T^\text{bulk}_{\mu\nu}\left(\phi\right)
\eeq
where $T^\text{bulk}_{\mu\nu}(\phi)=\partial_\mu \phi \partial_\nu \phi-\frac{1}{2}G_{\mu\nu}\partial_\alpha\phi\partial^\alpha\phi+\frac{1}{2}G_{\mu\nu}m_\Delta^2\phi^2$  is the leading order bulk stress tensor in $\mathcal{O}(1/N)$. We thus have for the integrand:
\beqn 
\langle O^\dagger(x_\psi^\star) T^\text{CFT}_{++}(x) O(x_\psi)\rangle &=&\int dz d^d x'\; \sqrt{G} 
D^{\mu\nu}_{++}(z,x';x) \left\lbrace \partial_\mu D^\phi(z,x';x_\psi )\partial_\nu D^\phi(z,x';x_\psi^\star) +...\right\rbrace\nonumber\\
&=& \int dz d^dy\;\sqrt{G} 
D^{\mu\nu}_{++}(z,x';x) T^\text{bulk}_{\mu\nu}\left(D^\phi(z,x'; x_\psi)\right)
\eeqn
where $D^{\mu\nu}_{\alpha\beta}(z,x';x)$ and $D^\phi(z,x';x_\psi)$ are the bulk-to-boundary  propagators for the graviton and scalar respectively. 
We identify the products of scalar boundary-to-bulk propagators as giving rise to the expectation value of the bulk stress tensor in the state $|\psi\rangle_\text{bulk}$:
\beq
T^\text{bulk}\left(D^\phi(z,x';x_\psi)\right) = \langle T^\text{bulk}_{\mu\nu}(z,x') \rangle_{\psi_\text{bulk}} 
\eeq
To see this, we focus on a particular term $\partial_\mu\phi \partial_\nu\phi$ in $T^\text{bulk}_{\mu\nu}(\phi)$. When viewed as a bulk operator, its expectation value in the bulk state $|\psi_\text{bulk}\rangle$ is given by:
\beq
\langle \partial_\mu \phi(z,x') \partial_\nu \phi(z,x')\rangle_{\psi_\text{bulk}}=\lim_{\epsilon,\epsilon' \to 0}\epsilon^{-\Delta}\epsilon'^{-\Delta}\langle \phi(\epsilon, x_\psi^\star)\partial_\mu \phi(z,x') \partial_\nu\phi(z,x')\phi(\epsilon,x_\psi)\rangle
\eeq
The leading order (disconnected) diagram of the bulk 4-point function is given by products of bulk-to-bulk propagator\footnote{ We analytically continue these propagators  to real time such that the ordering is the appropriate one for computing expectation values in the state $|\psi\rangle_\text{bulk}$. }:
\beqn
\langle \phi(\epsilon, x_\psi^\star)\partial_\mu \phi(z,x') \partial_\nu\phi(z,x')\phi(\epsilon,x_\psi)\rangle &=&\partial_\mu\langle \phi(\epsilon,x_\psi^\star)\phi(z,x')\rangle \partial_\nu\langle \phi(z,x')\phi(\epsilon',x_\psi)\rangle\nonumber\\
&=&\partial_\mu D^\phi_\text{bulk-to-bulk}(\epsilon,x_\psi^\star;z,x')\partial_\nu D^\phi_\text{bulk-to-bulk}(z,x';\epsilon',x_\psi)\nonumber
\eeqn 
The boundary-to-bulk and bulk-to-bulk propagators are related by the limit:
\beq
D^\phi(z,x';x)=\lim_{\epsilon\to 0}\epsilon^{-\Delta}D^\phi_\text{bulk-to-bulk}(z,x';\epsilon, x)
\eeq
We thus see that $\partial_\mu D^\phi(z,x';x_\psi^\star)\partial_\nu D^\phi(z,x';x_\psi) = \langle \partial_\mu \phi(z,x')\partial_\nu \phi(z,x')\rangle_{\psi_\text{bulk}}$. Similar relations hold for the other two terms in $T^\text{bulk}_{\mu\nu}$ and we conclude that: 
\beq
\langle O^\dagger(x_\psi^\star) T^\text{CFT}_{++}(x^+,x^i) O(x_\psi)\rangle = \int dz d^d x'\;\sqrt{G} 
D^{\mu\nu}_{++}(z,x';x)\langle T^\text{bulk}_{\mu\nu}(z,x')\rangle_{\psi_\text{bulk}}
\eeq 
We need to integrate this relation over $\int dx^+ d^{d-2}\vec{x} \zeta^+(x^i)$ on the boundary. In particular, since $\zeta^+(\vec{x})$ is independent of $x^+$, we can take it out of the null integral, and replace $\int dx^+ D^{\mu\nu}_{++}(z,x';x^+,\vec{x}) = \delta^{\mu\nu}_{++}D^\text{shock}(z,\vec{x}';\vec{x})\delta(x'^-)$, where $D^\text{shock}(z,\vec{x}';\vec{x})$ is the boundary-to-bulk propagator for the shock wave graviton mode: $h^{++}(z,x'^-,\vec{x})=f(z,\vec{x}')\delta(x'^-)$. In $\text{AdS}_{d+1}$, $D^\text{shock}(z,\vec{x}';\vec{x})$ is determined by solving Einstein's equations for this metric fluctuation giving
the shock-wave equation:\footnote{This shock wave metric is actually a full non-linear
solution to Einstein's equations although we have not used this fact. }
\beq\label{eq:shock}
\left(\partial^2_z + \partial^2_i-\frac{d+3}{z}\partial_z + \frac{2d+4}{z^2}\right) D^\text{shock}(z,y^i;x^i)=0,\;\; \lim_{\epsilon \to 0} D^\text{shock}(\epsilon,y^i;x^i)\to \epsilon^2 \delta^{d-2}(x^i-y^i)
\eeq 
The factor $\delta^{\mu\nu}_{++}\delta(x'^-)$ localizes the bulk integral onto $\mathcal{L}^+(\partial\mathcal{M}_0)$, and projects onto the $(++)$ component of bulk stress tensor:
\beqn\label{eq: CFT_mod_full_final}
\int_{\mathcal{L}^+(\partial\reg)}\zeta^+(x^i)\langle T^\text{CFT}_{++}(x^+,x^i) \rangle_{\psi_\text{CFT}} &=& \int_{\mathcal{L}^+(\partial\mathcal{M}_0)} \sqrt{h} \tilde{\zeta}^+(z,\vec{x}')\langle T^\text{bulk}_{++}(z,x'^-=0,x'^+,\vec{x}')\rangle_{\psi_\text{bulk}}\nonumber\\
\tilde{\zeta}^+(z,\vec{x}')&=& z^{-2}\int d^{d-2}\vec{x} \zeta^+(\vec{x})D^\text{shock}(z,\vec{x}';\vec{x})
\eeqn
One can finally check from (\ref{eq:shock}) that $\tilde{\zeta}^+(z,\vec{x}')$ satisfies the extremal bulk extension of $\zeta^+(\vec{x})$ on $\partial\mathcal{M}(\reg)$:
\beq\label{eq: AdS_extr}
\left(-\frac{d-1}{z} \partial_z +\partial^2_z+\partial_{\vec{x}'}^2\right) \tilde{\zeta}^+(z,\vec{x}')=0,\;\;\lim_{\epsilon\to 0}\tilde{\zeta}^+(\epsilon,\vec{x}') \to \zeta^+(\vec{x}')
\eeq
which is precisely what defines $\zeta^+_\text{bulk}(z,\vec{x})$, making (\ref{eq: CFT_mod_full_final}) equivalent to (\ref{eq: full_mod_equal}), consistent with JLMS formula.

\section{Discussion} 

In this paper, following the circle of ideas in \cite{Wall:2009wi,Wall:2011hj,Blanco:2013lea}, we established a relation between the monotonicity of relative entropy and the averaged null energy condition in arbitrary QFTs, and in so doing proved the most general Hofman-Maldacena bounds on the data in CFT three-point functions. We will now summarize the perturbative calculation we performed to establish this connection and then conclude with possible future work. 

The general goal was to study perturbatively the shape dependence of modular Hamiltonians/energies. We did this by applying ``perturbation  theory for reduced density matrices'' which turns out to have some novel features which we describe now. 
Schematically the important term in our calculation \eqref{dln} came from expanding the log
used to define the modular Hamiltonian.  Here we give an alternative description of this expansion: 
\beqn
\label{noncomm}
- \ln \rho_{A_0} ( 1 + \rho_{A_0}^{-1} \delta \rho )
&=&  K_{A_0} - \sum_{n=0}^{\infty}(-1)^n\frac{B_n}{n!} \underbrace{\left[ K_{A_0}, \left[ K_{A_0} , \ldots \left[ K_{A_0}, \rho_{A_0}^{-1} \delta \rho \right] \right] \right]}_{n-\mathrm{times}}+ O(\delta \rho^2)
\eeqn
where $B_n$ are the Bernoulli numbers. The right hand side comes about due to the non commutativity of the two matrices in the log on the left hand side. That is, these are the usual nested commutator terms in the Baker-Campbell-Hausdorff formula keeping only terms to order $\mathcal{O}(\delta \rho)$ (see also \cite{Kelly:2015mna} for related discussion). 

This set of nested commutators clearly has something to do with the evolution with respect
to $ K_{A_0}$ - or in other words modular flow. 
So it should come as no surprise that these terms can be re-summed into an integral over $\rho_{A_0}^{-is/2\pi }\left(\rho_{A_0}^{-1} \delta \rho\right) \rho_{A_0}^{is/2\pi}$ multiplied by some kernel - a fact we used in \eqref{dlnints}. In fact, in going from equation \eqref{noncomm} to \eqref{dlnints}, one simply uses the following integral representation of the Bernoulli numbers \cite{NIST:DLMF, Olver:2010:NHMF}:\footnote{Note that we pick the convention where $B_1 = +\frac{1}{2}$; also recall that $B_{2m+1} = 0$ for $m=1,2\cdots $. The corresponding terms in the integral representation \eqref{BN} pick out the residue at $s=0$, which is only non-trivial for $n=1$.}
\beq \label{BN}
B_{n} = -\frac{(-i)^{n}}{(2\pi)^{n}}\int_{-\infty}^{\infty}ds\;\frac{s^{n}}{4\,\mathrm{sinh}^2(\frac{s+i\epsilon}{2})} \;\;\; \cdots \;\;\;(n\in \mathbb{Z}).
\eeq
Surprisingly this integral and kernel as well have the effect of switching the original Euclidean diffeomorphism contained in $\delta \rho$ and used to move around the entangling surface, to a real time diffeomorphism determined by the new vector field given in \eqref{realb}.  From here the null energy operators involved in the ANEC just pop out as boundary terms when integrating by parts over the real time diffeomorphism. Of course in real times now a new boundary has opened up; what previously was the co-dimension 2 entangling surface  at the origin in Euclidean space becomes a null hypersurface along the Rindler horizon where the null energy operators are defined. 

The non-commutativity emphasized in \eqref{noncomm} was of fundamental importance to our calculation. We feel that we do not fully understand the magic behind this calculation and that there are new surprises lurking if we go to higher orders in perturbation theory and try to systematize this approach. 
Similar tools were applied in \cite{Faulkner:2015csl,Faulkner:2014jva} to entanglement entropy where it was important to control these commutator terms
in order to find agreement between this perturbative approach and known results from AdS/CFT.
Here we have also established a similar agreement with AdS/CFT and in particular the recent proposal by JLMS \cite{Jafferis:2015del} for the modular Hamiltonian in AdS/CFT. 

Apart from gaining a deeper understanding into the inner working of these calculations we now give some detail of future work that we think would be valuable to pursue. 

\subsection{Sharpening the argument}

In the main sections of the paper our derivation eschewed any issues related to the precise definition
of entanglement and modular energy in quantum field theory.  Indeed these quantities
are expected to be afflicted by significant UV divergences, and possibly even ambiguities related
to how one  splits the degrees of freedom between $A$ and $A^c$. Thus in order
to calculate these quantities we must specify a regulator and a prescription for splitting the Hilbert space. However it became clear to us that we never needed to do this, and so any real discussion of a regulator was relegated to Appendix~\ref{app:cutoff}.

Ultimately this should not have come as a surprise, the final goal was to calculate either relative
entropy or the full modular Hamiltonian - both of which are expected to be UV finite quantities
and both of which can actually be given a definition directly in the continuum \cite{araki1976relative,haag2012local}. This definition
however was not convenient for our current calculation so
at an intermediate step we needed to calculate the expectation of the full modular Hamiltonian
in terms of the UV sensitive (half) modular Hamiltonian. Since we never explicitly saw these UV divergences, our manipulations should be regarded as formal.\footnote{They might be regarded as about as formal as the usual derivation of the replica trick for Renyi entropies in terms of a partition function on a singular surface \cite{Calabrese:2004eu}. }  Appendix~\ref{app:cutoff} is an attempt to remedy this, by giving some details
of a brick wall like regulator \cite{hooft1985quantum} that renders the modular energy and associated quantities
well-defined. The brick wall regulator introduces dependence on the boundary conditions
one chooses for fields at the wall close to the entangling surface. 

The regulated version of relative entropy does not satisfy the property of monotonicity (for a finite but small cutoff scale) since the brick wall cutoff is a rather drastic modification to the theory that does not allow one to compare different spatial regions with the same modification. So to claim a completely rigorous proof of the ANEC  we still need to show that when we remove the brick wall cutoff the quantity we get is the continuum version of relative entropy - which is then known to be monotonic \cite{araki1976relative}.  This requires methods that are beyond the scope of this paper, and we leave this to future investigations. Ultimately we would like a 
mathematically rigorous derivation, perhaps without reference to density matrices and using methods of algebraic quantum field theory \cite{haag2012local, zbMATH00845667}.  

Finally we would like to understand if there are any restrictions on the state in which we calculate the expectation value of the deformed modular Hamiltonian. For example we formulated our state in terms of a local operator insertion at $x_E^0 = \pm \tau$, which is sufficiently general for a CFT. More generally, say for relativistic theories, our argument will go through relatively unmodified if we just insert a general state of the theory and its conjugate in flat space along the Euclidean time slices $x^0_E = \pm \tau$. However we are required to separate the diffeomorphism that moves around the entangling surface away from $|x^0_E| \geq \tau$. We can make the region in which the diffeomorphism acts small but we should be limited by $|\zeta|$ the size of the diffeomorphism vector field at the entangling surface. This presumably puts some restriction on the state such that the expectation value of the stress tensor cannot get arbitrarily large. For example if we work with the state created by a local operator insertion  $| \int_{\mathcal{H}_\pm}  \left< T_{\pm\pm} \right>_\psi | \sim \tau^{-1} < |\zeta|^{-1} $. This is likely just the restriction that the perturbative expansion  converge and we can always arrange this to happen by taking a small enough spatial deformation.

 \subsection{Generalizations}

One obvious generalization involves attempting to prove the ANEC in other space-times as well as along more general complete achronal null geodesics.\footnote{These are geodesics where no two points on the curve are timelike separated. The ANEC is known to fail in curved space-times where the null geodesic is chronal \cite{Graham:2007va,visser1996gravitational}. }
Along these lines it might be an easier first step to try to apply the methods of this paper
to stationary but not static black holes with the null generator lying along a bifurcate Killing horizon (like the Kerr black hole). Since we used the framework of perturbation theory starting from a state described by a known density matrix (the Hartle-Hawking state) we are not very optimistic this will succeed when we don't have such a starting point. 

 Instead perhaps a more fruitful direction to pursue would be to consider the generalized second law (GSL) for quantum fields outside of a black hole with a static bifurcate Killing horizon. 
Here we are referring to the work of \cite{Wall:2011hj}  where the GSL was proven for free as well as super renormalizable QFTs.\footnote{The Hawking area theorem proves
the GSL when the area term dominates in the $G_N \rightarrow 0$ limit - that is for a classical dynamical background where the classical matter satisfies the NEC.  As discussed in \cite{Wall:2011hj}, what remains, is to prove the GSL when classically the area does not increase - for quantum fields on a stationary black hole background plus free gravitons. For obvious reasons here we then focus on the static case, and leave out gravitons for simplicity. }
The GSL applies to the following generalized entropy:
\beq
S_{\rm gen} = \frac{ {\rm Area}(\partial A )}{4 G_N} + S_{EE}(\rho^\psi_A)
\eeq
where ${\rm Area}(\partial A)$ refers to the area of a codimension-1 slice of the Killing horizon where the spatial region $A$ ends ($\partial A$) and $S_{EE}$ is the entanglement entropy of the quantum fields outside this horizon slice. 
Applying the monotonicty of relative entropy to $S_{EE}(\rho^\psi_A)$ one finds:
\beq
\Delta S_{\rm gen}  \geq \frac{\Delta {\rm Area}( \partial A)}{4 G_N}  + \Delta \left< K_{A} \right>_\psi
\eeq
where now $ \Delta $ is  a \emph{finite} null deformation ($\Delta x^+ = \zeta^+(\vec{y}) $) of the entangling surface $\partial A$ to the future of the bifurcation surface $\partial A_0$. The change in the area is
simply due to the perturbative back reaction of the quantum fields on the space-time via Einstein's equations:
\beq
\label{gsl}
\Delta S_{\rm gen}  \geq  2\pi \int d^{d-2}\vec{x} \left(  -  \int_{\zeta^+}^\infty dx^+ (x^+ - \zeta^+)\left< T_{++}\right>_\psi + \int_{0}^\infty dx^+ x^+\left< T_{++}\right>_\psi \right)  + \Delta \left< K_{A} \right>_\psi
\eeq
where we have made use of the Raychaudhuri equation with the correct future boundary condition appropriate for a causal horizon. 

To make further progress we need some handle on $K_{A}$ for general null deformations away from $A_0$. This does not sound very promising for our perturbative approach, however it does seem like we can carry out our calculations to arbitrary orders in $ \zeta^+$ \cite{allorders}.
 Thus with some luck we might be able to prove a statement about $K_{A}$ and get a handle on \eqref{gsl} and possibly
show the GSL in this case, $\Delta S_{gen} \geq 0$. A further hint comes actually from AdS/CFT. For a Rindler space cut we have carried out a more detailed calculation\footnote{This calculation has some overlap with \cite{Bunting:2015sfa} and the details will be reported elsewhere.} than that outlined in Section~4 where we previously showed the equality between the null energy operators in the bulk and boundary.
More generally one can  show for \emph{finite} null deformations, but small perturbations to the state (in the $1/N$ sense):
\begin{align} \nonumber
2\pi \int d^{d-2} \vec{x} \int_{\zeta^+}^\infty dx^+ \left(x^+ - \zeta^+ \right)
\left< T_{++}^{CFT} \right>_\psi & = \frac{{\rm Area}_{\partial \mathcal{M} \backslash A}(\delta g)}{4G_N} \\
& + 2\pi \int d^{d-2} \vec{x} \sqrt{h} \int_{\zeta^+_{\rm bulk}}^\infty dx^+ \left(x^+ - \zeta^+_{\rm bulk} \right)
\left< T_{++}^{\rm matter} \right>_\psi
\end{align}
and our notation is the same as that in  Section 4, where for example $\zeta^+_{\rm bulk}$ is the bulk HRT extremal surface corresponding to the deformation $\zeta^+$ on the boundary and $\mathcal{M}$ is the spatial region
between this extremal surface and $A$ on the boundary. Here the area term
is the change in the area of the extremal surface due to the backreaction on the metric $\delta g$
in the state $\psi$ (via Einstein's equations.)  Note that the extremal surface condition in pure AdS
for finite null deformations remains a linear differential equation that matches with the
infinitesimal version \eqref{eq: AdS_extr} and so $\zeta^+_{\rm bulk}$ is the same extension
as that used in Section~\ref{sec:jlms}. Now comparing this statement with
that of JLMS \cite{Jafferis:2015del} we could consistently identify the modular Hamiltonian for finite null
deformed regions as:
\beq
\label{guess}
K^{CFT}_{A} \mathop{=}^? 2\pi \int d^{d-2} \vec{x} \int_{\zeta^+}^\infty dx^+ \left(x^+ - \zeta^+ \right)
 T_{++}^{CFT} 
 \eeq
up to an additive constant, with a similar equation holding for the bulk region modular Hamiltonian
$K_{\mathcal{M}}^{ \text{bulk}}$.
This is certainly not a proof. We have made two guesses (for the bulk and the boundary) and shown them to be self-consistent. And in particular this only works for a special class of states that don't have a large back reaction on the bulk. Note that this last issue also plagued our comparison between the bulk and boundary for small deformations. 
We simply note here that our perturbative approach, when considered at higher orders, can possibly prove such a statement.\footnote{Actually a simpler argument is to  take the perturbative result we have derived
for null shape deformations and then use the QFT boost generator around the original undeformed Rindler cut to amplify the deformation. 
This boost will then act on the state. However if we work in a sufficiently general state this should not matter. This process seems to work, and agrees with \eqref{guess}, when trying to construct the full modular Hamiltonian and we leave the details of how this works for the half space modular Hamiltonian for the future. We thank Aron Wall for suggesting this argument to us.  } 
Of course if \eqref{guess} is true then the GSL follows trivially since the right hand side of the inequality
in \eqref{gsl} just vanishes.

Finally we point out that in some sense these calculations have  already been pushed  to higher orders.
Rather than consider the excited state modular energy, if we just calculate the modular Hamiltonian in the original vacuum state it should reproduce the entropy of the vacuum. At first order this vanishes but the second order variation
of entropy in a CFT was calculated in \cite{Faulkner:2015csl} using similar methods to this paper.
This quantity is sometimes referred to as entanglement density \cite{Bhattacharya:2014vja}. 
 Although it was not realized at the time the answer in \cite{Faulkner:2015csl} can be related to a correlation function of two  ``null energy operators'' - the same null energy operators that appear in the (half sided) modular Hamiltonian in this work. This will be the subject of a forthcoming paper \cite{edensity}. Taken together this hints at a unifying picture for vacuum entanglement in CFTs related to null energy operators that may even pave the way to a new understanding and proof of the Ryu-Takayanagi \cite{Ryu:2006bv} and HRT  \cite{Hubeny:2007xt} proposals for calculating entanglement entropy in the vacuum state of holographic CFTs.

\subsection*{Acknowledgements}
It is a pleasure to thank  Xi Dong, Gary Horowitz, Veronika Hubeny, Aitor Lewkowycz, Don Marolf,   Mukund Rangamani, David Simmons-Duffin and Aron Wall for discussions and suggestions.  Work supported in part by the U.S. Department of Energy contract DE-FG02-13ER42001 and DARPA YFA contract D15AP00108.

 \appendix

\section{Cutoff at the entangling surface}
\label{app:cutoff}

In this Appendix we would like to give a prescription for regulating the modular energy that
we calculate in the main part of the paper. We go through this in some depth because  the arguments we gave previously were somewhat formal. Although the quantity in which we are ultimately interested -- the full modular Hamiltonian --  is UV finite \cite{haag2012local}, at intermediate steps we encountered quantities which are not.  In particular the modular energy of some state is expected to have the same UV divergences as the entanglement entropy of that state because the difference between them is the relative entropy which is UV finite.\footnote{There are still several reasons to expect some of our intermediate steps to be finite. Any divergences should be local to the entangling surface, and assuming our regulator is \emph{geometric}\cite{casini2015mutual, liu2013refinement,grover2011entanglement} no such term which respects the $S(A)=S(A^c)$ purity condition can generate a divergence for first order \emph{spatial} shape deformations. Similarly there is a general expectation that any such divergences  cancel in the difference $S(\rho^\psi_A) - S(\rho_A)$ although we will find evidence that this cancellation might not always occur. Of course these variations and differences are still calculated in terms of divergent quantities so we proceed. } Thus the issues here are the same as the usual issues of defining entanglement entropy in the continuum \footnote{For a recent
discussion of some of the issues involved see \cite{ohmori2015physics, casini2015mutual}. When the QFT in question is a gauge theory there are even questions about how the degrees of freedom are split between two spatial regions \cite{casini2014remarks}.}. There are several ways to define a regulated version of entanglement entropy, but the most convenient for us will be a ``brick wall'' regulator \cite{hooft1985quantum}. This is so we can still use Euclidean path integral methods to construct the density matrices in question. Apart from possible IR issues the entropies are now finite - the IR issues do not concern us and cancel when evaluating the differences between excited and vacuum states, at least for states that are sufficiently close to the vacuum near the boundaries of space. 

\myfig{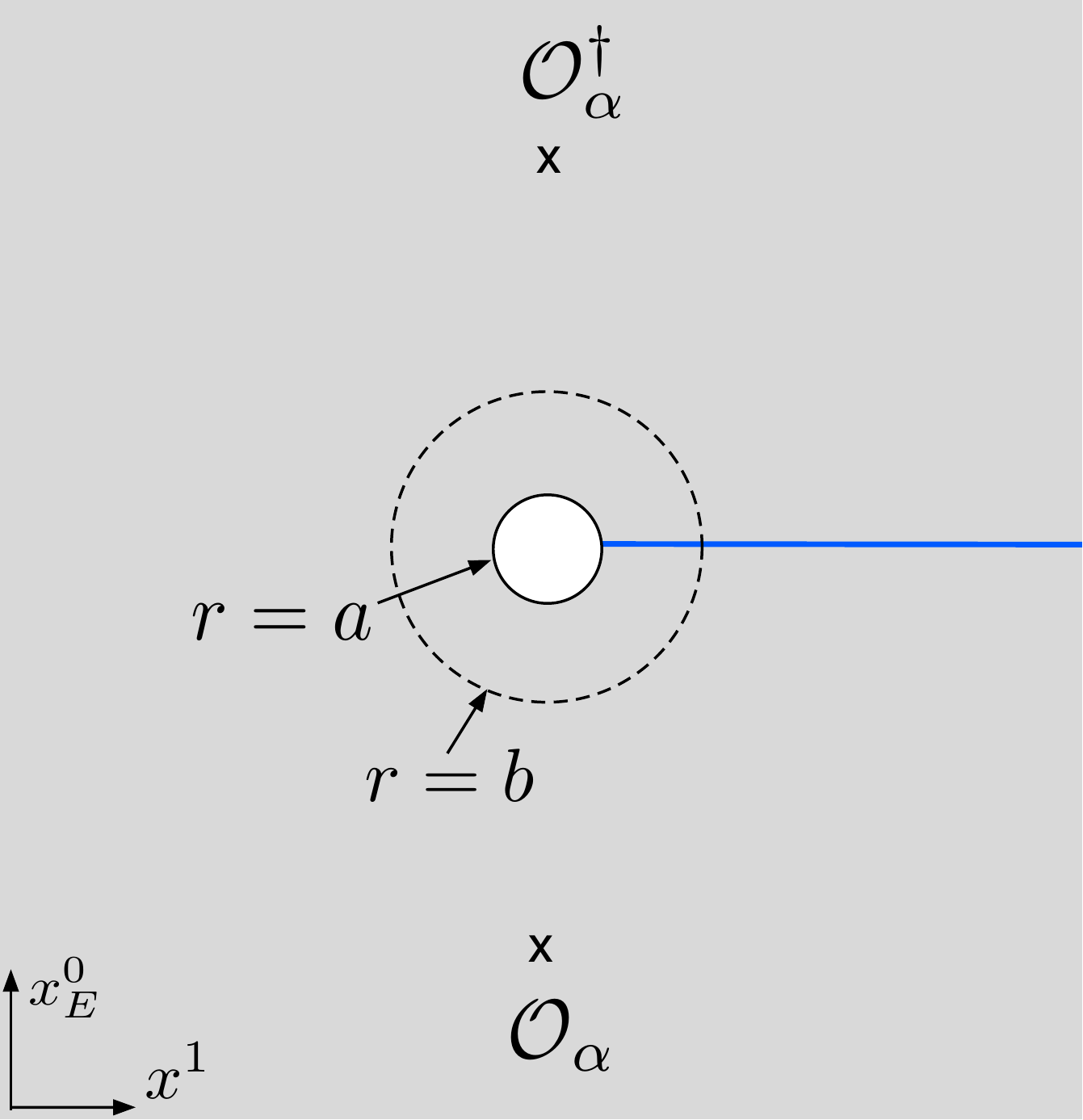}{6}{\small{\textsf{The path integral construction of the regulated reduced density matrix for an excited state. We cut out a cylindrical region of radius $r=a$ around the entangling surface, with brick-wall-like boundary conditions. Also shown is the fictitious cutoff surface of radius $r=b$.  }} }

Roughly speaking we can simply go through our calculation in Sections~\ref{sec:moddef} and \ref{subsec:comp} with
the reduced density matrices defined via path integrals on manifolds with a cylindrical region of radius $a$ cut out from around the entangling surface: $\rho_{A_0, g} \rightarrow \rho_{A_0, g}(a)$ (see figure \ref{fig:fig2.pdf}).  In order to to do this consistently we should impose boundary conditions on the cutoff surface - we will assume that the boundary conditions at the cutoff surface decouple in the limit $a\to 0$. We might also need to add new degrees of freedom here \cite{donnelly2015entanglement,donnelly2015geometric} and there are good reasons to believe these should
decouple when calculating such things as relative entropy \cite{casini2014remarks}. We also require the following:
\begin{itemize}
\item Rotation/Boost invariance for the undeformed Rindler region. This is so that $K_{A_0}$ still has the interpretation as the generator of rotations/boosts around the cutoff surface. For example this
will require that the stress tensor at the cutoff surface is constrained to have zero
rotation flux $T_{\theta r}|_{r \rightarrow a} \rightarrow 0$ into the cutoff cylinder. This should be required as part of the boundary conditions.\footnote{Note however that this can fail in the case of chiral theories, in which case the boost symmetry is anomalous \cite{Castro:2014tta, Iqbal:2015vka, Nishioka:2015uka, Hughes:2015ora}.} 

\item For a more general region - we cut out a cylinder in Gaussian normal coordinates. Here of course we do not have rotation invariance. 
We use normal coordinates so we can still use the relation \eqref{dm} derived in the main text. One way to do this is to pick the diffeomorphism to map the deformed entangling surface to Gaussian normal coordinates - where the regulator is then picked to be a metric distance $a$ orthogonal to the surface away from $A$. For us this amounts to the choices:
\begin{align}
\zeta &= \zeta^z \partial_z +\zeta^{\bar{z}} \partial_{\bar{z}} +\frac{1}{2} \partial^i(\zeta^z)  \bar{z} \partial_i  
+\frac{1}{2} \partial^i(\zeta^{\bar{z}})  z \partial_i   + \ldots \\
g_{\mu\nu} dx^\mu dx^\nu  &= - dzd\bar{z} + \left( \delta_{ij} + \bar{z} \partial_i \partial_j \zeta^z + z \partial_i \partial_j \zeta^{\bar{z}}    \right) d x^i dx^j  + \ldots
\end{align}
where we have expanded the diffeomorphism and the metric close to the entangling surface. 
We then cutoff the path integral which defines $\rho_{A_0, g}(a)$ in \eqref{dm}  at $r = |z| = a$ supplying some appropriate yet unspecified boundary conditions.  After making this slight modifications the diffeomorphism acts the same way as in the bulk of the text - in particular there is no boundary term due to the displacement of the cutoff surface (although of course the new stress tensor could have delta function contributions on the cutoff surface.)

\item At the very minimum we require that for some local operator inserted in the path
integral that defined $\rho_{A_0, g}(a)$ we should have:
\beq
\label{req}
\lim_{a \rightarrow 0} {\rm Tr}_{A_0} \rho_{A_0, g}^{\psi}(a) \mathcal{O}(x) =  \left<  \mathcal{O}(x)  \right>_\psi
\eeq

\end{itemize}
 
Following the steps below \eqref{tospec}  in Section~\ref{subsec:modham} for the change in modular Hamiltonian
due to the stress tensor deformation, the differences are due 
to a slightly modified diffeomorphism and a different region of integration for the stress tensor  $R_0$ in the Euclidean plane which is cutoff by $r > a$; see figure \ref{fig:fig2.pdf}. This cutoff is distinct from the \emph{imaginary cutoff surface} defined in Section~\ref{subsec:modham} with $r >b$ and we will take $ a \ll b$.  Indeed splitting the contribution from the stress tensor integral into the
three regions as we did previously, there is only one term which is sensitive to the cutoff in the limit $a \ll b$ and this is the contribution from \emph{inside the hole} $ a < r< b$ which we call $\delta_\zeta K_b(a)$. The other two contributions from the \emph{branch cut} and the  \emph{imaginary cutoff surface} give identical results as in the main text.
With the same set of manipulations we can write the potentially problematic term as:
\beq
\label{holenew}
\delta_\zeta K_b(a) = 2 \pi \int_{a < r< b} d^{d}x :T_{\mu\nu}: \delta  \widetilde{g}^{\mu\nu}
\eeq
where the resulting real time metric deformation is:
\beq
\delta \widetilde{g}_{\mu\nu} dx^\mu dx^\nu = \frac{1}{2} \left( \Theta(x^0)  x^- \partial_{i} \partial_j \zeta^+ +  \Theta(- x^0)  x^+ \partial_{i} \partial_j \zeta^- 
  \right) dx^i d x^j
\eeq
Note the integration region is now a section of a solid hyperboloid. This is slightly modified from \eqref{realb} and \eqref{realb2} since we are now working in Gaussian normal coordinates. Of course to analyze
the limiting properties of \eqref{holenew} we should  trace it against our state ${\rm Tr} \rho_{A_0}^\psi \left( \cdot \right)$.

At this point it is possible to remove the brick wall regulator $a \rightarrow 0$ from \eqref{holenew}. Dependence on $a$ appears in the integration region $R_0$ as well as implicitly in $T_{\mu\nu}$ since this
is the appropriate field theory stress tensor in the presence of a boundary. Note that
the boundary conditions on the brick wall in Euclidean space have naturally been mapped to Rindler space in real times along the hyperbola $r=a$, $ - \infty  < s < \infty$.
The claim is that we can remove the regulator from the integrand using the requirement
\eqref{req}. Of course the remaining  $ d^d x$ integral may still be divergent, however we found this not to be the case in the main text.   It is this sense in which we expect the boundary conditions on the surface $r=a$ to decouple. To show this rigorously we would have to show the integrand converges sufficiently
 uniformly to the $a=0$  limit. Note that the metric deformation  $\delta g_{ij} \sim r e^{-|s|}$ in Rindler
 coordinates  and so this is a mild condition on the behavior of the stress tensor close to the brick wall.\footnote{Note because of the $::$ vacuum subtraction for the stress tensor any divergence that might appear exactly
 at the cutoff surface when $a$ is fixed (say due to an image charge) is state independent and will cancel. The potential divergence we are worried about is in the subsequent limit $a \rightarrow 0$. }
 To say anything further  we would need to specify more about the boundary conditions than we are willing to. However since the details of the boundary conditions are not  important for defining relative entropy or the full modular Hamiltonian in the continuum of a QFT, it must be the case that any divergences we might see here should cancel when calculating these final quantities. Instead we can turn this condition around and demand that this should be true for any brick wall regulator that is supposed to be a good regulator for calculating modular energy.

We now turn to the second contribution to the deformed modular energy (the first term in \eqref{t1}.)
Compared to our previously obtained expressions we now find a contribution from
the boundary of the cutoff region at $r = a$ which looks like:

\beq\label{cut1new}
\left.\mathrm{Tr}_{\reg}\left(\delta_{\zeta}\rho_{\reg}^{\psi}K_{\reg}\right)\right|_{\pa R_a} =  a\oint_{\partial R_a}\zeta_{\mu}n_{\nu}\Big(\left\langle T^{\mu\nu}(x)K_\reg \right\rangle_{\psi}-\langle T^{\mu\nu}(x)\rangle_{\psi}\langle K_\reg \rangle_{\psi}\Big)
\eeq

If we instead calculate this contribution to the shape deformation of the full modular hamiltonian (this was defined as  $\mathfrak{T}^{(1)}$ in \eqref{t1}) we get a term coming from the complement $\regc$ which adds to give the total contribution to $\mathfrak{T}^{(1)}$ coming from the cutoff surface $\pa R_a$:
\beq
\label{ttstress}
\left. \mathfrak{T}^{(1)} \right|_{\pa R_a} = - a\oint_{\partial R_a}\zeta_{\mu}n_{\nu}\Big(\left\langle T^{\mu\nu}(x)\cK_\reg \right\rangle_{\psi}-\langle T^{\mu\nu}(x)\rangle_{\psi}\langle \cK_\reg \rangle_{\psi}\Big)
\eeq
where we remind the reader that $\widehat{K}_{A_0}$ is the undeformed \emph{full} modular Hamiltonian. 
In the limit $a\to 0$, the above term appears to be linearly suppressed; however one might worry that there are potential enhancements from the stress tensor coming close to $\cK_{\reg}$ in the first term above. To see that this does not happen, recall that $\cK_\reg$ is a conserved charge, namely the generator of rotations around the entangling surface. Consequently, we can freely move it away from cutoff surface as well as the other stress tensor inside the above correlator.  Here we have to take into account the fact that the boundary condition on the cutoff surface should not allow for any $\widehat{K}_{A_0}$ flux into the cutoff surface $T_{r\theta} \rightarrow 0$. If we could move $\cK_\reg$ off to $x^0_E = \pm \infty$, then the corresponding term would vanish, because $\cK_\reg$ annihilates the vacuum. However, as we keep moving $\cK_{\reg}$ away from the stress tensor, we will eventually cross the operator insertions $\cO_{\a}$ or $\cO_{\a}^{\dagger}$ (depending on whether we move $\cK_{\reg}$ towards $x_E^0 \to -\infty$ or $x_E^0\to +\infty$). Every such crossing gives a non-trivial contribution of the form $\langle T^{\mu\nu} [\cK_{\reg}, \cO_{\a_m}]\cdots\rangle$, where $\cdots$ denotes the remaining operator insertions. 
 However, it should now be clear that these remaining terms are correlation functions between well-separated operators (as long as $\tau \gg a$), and we do not get any enhancement to cancel the factor of $a$. Therefore, we conclude that the contribution from the cutoff surface to $\mathfrak{T}^{(1)}$ vanishes in the limit $a\to 0$. We claim victory.
 
 Before moving on we note that if we did not do the subtraction that defined the full modular energy, the term \eqref{cut1new} might still be divergent. We can give the following crude estimate for any such divergence.  Note that the half sided modular Hamiltonian, as an integral over the stress tensor, can still be moved around but now it is always anchored to the cutoff surface. We can use this freedom to move the two stress tensors in \eqref{ttstress} as far apart as possible - on opposite sides of the hole. To get an estimate we now replace the correlation function  in the first term of \eqref{cut1new} with the CFT correlation function without the cutoff surface - on flat Euclidean space. 
We need to consider the OPE of two stress tensors schematically of the from:
 \beq
 T T \sim  \sum_{k,\alpha} C_{\alpha} (\delta x^2)^{- d+ \Delta_\alpha/2 +k/2 }   \partial^k \mathcal{O}_\alpha
 \eeq
 where $\delta x^2 \sim (a+r)^2 + (\vec{x} - \vec{x}')^2$ and where $r >a$ refers to the location of
 the modular Hamiltonian stress tensor. Here $\mathcal{O}_\alpha$ are some local primary operators and only scalars can possibly contribute a divergence. Close to the entangling surface for any divergent term we
 can expand $ \zeta^{\pm}(\vec{x})$ and $ \left< \mathcal{O}_\alpha (\vec{x}') \right>_\psi$ by taking $\vec{x}' \sim \vec{x}$. We get some leading term
 from the unit operator. But there are good reasons this term should vanish. Firstly it is state independent and so should occur for  the vacuum state, but there is simply no term
 we can write down at linear order in $\zeta$ which is local to the entangling surface and has the required rotation/boost invariance around the entangling surface. However now consider a non unit operator, we no longer have rotation around the entangling surface. Then using scale invariance we only expect a divergent contribution to the modular energy of the form:
\begin{align}
C_{\alpha} a \int d^{d-2}\vec{x} & \int d^{d-2}\vec{x'} \int_a d r r  (\delta x^2)^{-d+\Delta/2 +1/2}\left<\partial_{\pm} \mathcal{O}_\alpha (\vec{x}') \right>_\psi  \zeta^{\pm}(\vec{x}) \\
&\sim a^{2-d+\Delta_\alpha}  C_{\alpha} \int d^{d-2} \vec{x} \left<\partial_{\pm} \mathcal{O}_\alpha(\vec{x}) \right>_\psi  \zeta^\pm(\vec{x})  \label{newdivdiff}
\end{align}
Naively one would have expected that such a contribution is not possible for uniform
deformations -  $\zeta^{\pm}$  independent of $\vec{x}$ - since in that case we can
write an expression for $K_{A}$ and $K_{A_0}$ in the absence of the cutoff surface and there is seemingly no divergence. However since these are half sided modular Hamiltonians it seems we should have allowed for the possibility that even the un-deformed $K_{A_0}$ has a local divergence:
\beq
K_{A_0} \sim  a^{2-d+\Delta_\alpha}  C_{\alpha} \int_{\partial A_0} \mathcal{O}_\alpha + {\rm finite}
\label{newdiv}
\eeq
At least this seems to be required if we want the answer to be consistent with our UV regulator and diffeomorphism invariance. This calculation is far too crude to be trusted, but it does suggest
that any kind of brick wall cutoff leaves one susceptible to state dependent divergences in the half sided modular energy of the above nature. In the main text we could have taken such divergences into account simply by adding \eqref{newdiv} and this would have generated the term in \eqref{newdivdiff} without any need for a brick wall.   

This new divergent contribution occurs if there is a very relevant $\Delta_\alpha \leq d-2$ scalar
operator appearing in the $TT$ OPE. For example it cannot be charged under any symmetries. Symmetries would also disallow \eqref{newdiv}.  It is not clear a theory with such a scalar operator can exist - see \cite{Bousso:2014uxa} for a related appearance of such operators in state dependent divergences for entanglement entropy. In recent proofs of the Hofman-Maldacena bounds from bootstrap methods, these operators also make an appearance \cite{Hartman:2016dxc}. In our work they are always harmless and cancel when we calculate the full modular Hamiltonian. 


\providecommand{\href}[2]{#2}\begingroup\raggedright\endgroup

\end{document}